\documentclass{ecai} 

\usepackage{xr-hyper} 
\usepackage{latexsym}
\usepackage{tabularx}
\usepackage{amssymb}
\usepackage{amsmath}
\usepackage{amsthm}
\usepackage{booktabs}
\usepackage{enumitem}
\usepackage{graphicx}
\usepackage{color}

\usepackage{graphicx}
\usepackage{subcaption}
\usepackage{tablefootnote}
\usepackage{float}
\usepackage{algorithm}
\usepackage{algpseudocode}
\usepackage{wrapfig}
\usepackage{array}  
\usepackage{booktabs}
\newcommand{\ra}[1]{\renewcommand{\arraystretch}{#1}}
\usepackage{multirow}
\usepackage{pdfpages}
\usepackage{xspace}
\usepackage{cite}
\usepackage{hyperref}

\newcommand{\name}{\textsc{Retro-li}\xspace}

\begin{document}
\begin{frontmatter}


\paperid{2402} 


\title{\name: Small-Scale Retrieval Augmented Generation Supporting Noisy Similarity Searches and  Domain Shift Generalization}


\author[A,B]{\fnms{Gentiana}~\snm{Rashiti}
}
\author[A]{\fnms{Geethan}~\snm{Karunaratne}
}
\author[B]{\fnms{Mrinmaya}~\snm{Sachan}} 
\author[A]{\fnms{Abu}~\snm{Sebastian}} 
\author[A]{\fnms{Abbas}~\snm{Rahimi}\thanks{Corresponding Author's Email: abr@zurich.ibm.com.\\ 
Published as a conference paper at ECAI 2024.}} 

\address[A]{IBM Research -- Zurich} 
\address[B]{ETH Zürich}

\begin{abstract}
The retrieval augmented generation (RAG) system such as \textsc{Retro} has been shown to improve language modeling capabilities and reduce toxicity and hallucinations by retrieving from a database of non-parametric memory containing trillions of entries. We introduce \name that shows retrieval can also help using a small scale database, but it demands more accurate and better neighbors when searching in a smaller hence \emph{sparser} non-parametric memory. This can be met by using a proper semantic similarity search. We further propose adding a regularization to the non-parametric memory for the first time: it significantly reduces perplexity when the neighbor search operations are noisy during inference, and it improves generalization when a domain shift occurs. We also show that the \name's non-parametric memory can potentially be implemented on analog in-memory computing hardware, exhibiting $O(1)$ search time while causing noise in retrieving neighbors, with minimal (<1\%) performance loss. Our code is available at: \href{https://github.com/IBM/Retrieval-Enhanced-Transformer-Little}{https://github.com/IBM/Retrieval-Enhanced-Transformer-Little} 
\end{abstract}

\end{frontmatter}


\section{Introduction}
\label{sec:intro}
Natural language processing is a rapidly growing field in machine learning. Advancements in large language models are mostly a product of the neural scaling law~\citep{scalinglaw}. 
Not only are language models themselves becoming larger, but their training data also follows suit. While \mbox{GPT-3~\citep{gpt3}} was trained on 300 billion tokens, \mbox{GPT-4~\citep{gpt4}} was trained on around 13 trillion tokens. As a result, the information retained by a language model is difficult to update or fact-check. Many of these data sources contain discriminatory language or misinformation, which are difficult to remove given the sizes of these datasets~\citep{gehman-etal-2020-realtoxicityprompts}.

\begin{figure}[t]
    \centering
     \resizebox{5cm}{!}{
    \begin{subfigure}[t]{0.22\textwidth}
    \includegraphics[]{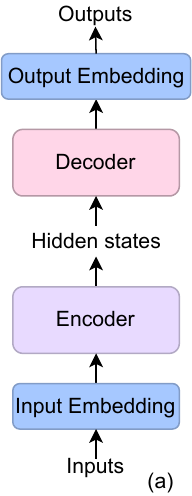}
    \end{subfigure}
    \begin{subfigure}[t]{0.2\textwidth}
    \includegraphics[]{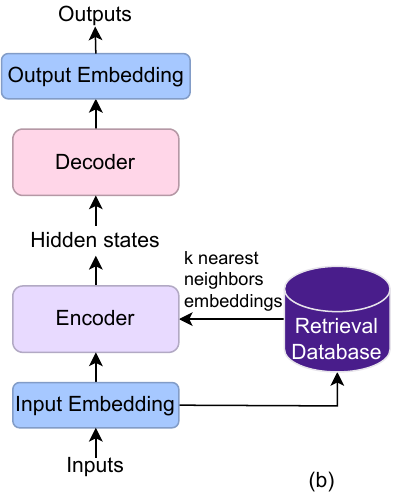}
    \end{subfigure}
     }
    \caption{(a) Classic encoder-decoder  (b) RAG encoder-decoder.}
    \label{fig:rag-system-simplified}
\vspace{0.5cm}
\end{figure}

In retrieval augmented generation (RAG)~\citep{rag}, a language model is enhanced with a non-parametric memory as depicted in Figure~\ref{fig:rag-system-simplified}. During training and inference, for each input sequence, RAG searches this memory for the $k$ most similar sequences (the so-called nearest neighbors) and uses them as additional input. RAG has been shown to help with under-fitted language models as the non-parametric memory increases with the number of training tokens~\citep{chinchilla}. Moreover, it has been shown to reduce toxic language and hallucinations~\citep{nvidiaretro} through its use of retrieval. Furthermore, unlike in purely parametric models, it is straightforward and does not require extensive computing power to update the retrieval database of a RAG model to reflect updated information. Finally, in a question-answering context, the ability to provide sources for the answer helps combat misinformation. This provides users with the opportunity to double-check the responses themselves, aiding the models' interpretability.

Despite all the aforementioned functional advantages of RAG, in many cases, the retrieval database of non-parametric memory contains trillions of tokens to be searched. In fact, in \textsc{Retro}~\citep{retro}, the scale begins at a retrieval database size of billions of tokens, which raises the question of search speed and optimization. Although similarity search acceleration libraries such as \textsc{Faiss}~\citep{faiss} and ScaNN~\citep{scann} can compare millions of vectors in milliseconds, they quickly become a bottleneck, especially for retrieval databases containing trillions of tokens 
(e.g., see Section~\ref{sec:hardware-aware_regularization})
. To address this critical bottleneck, we suggest three directions whose effectiveness is shown in our proposed \name.

The contributions of this paper are as follows:

\begin{figure*}
\centering
\includegraphics[width=0.9\linewidth]{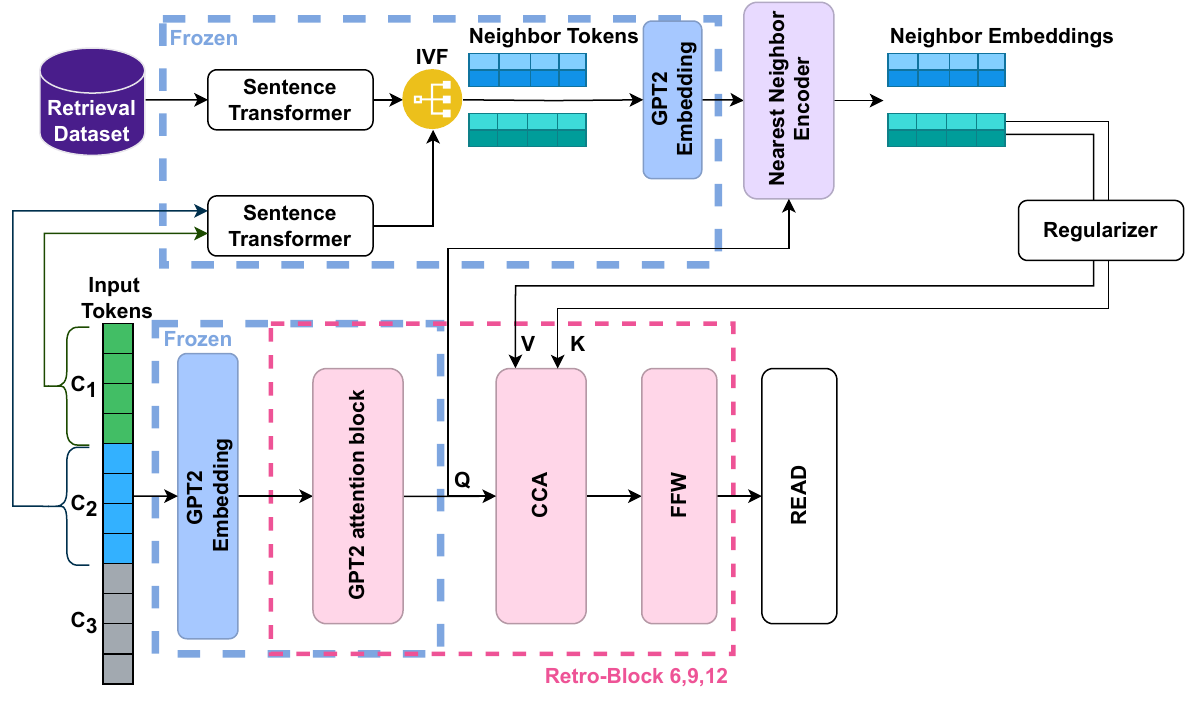}
\caption{\name architecture diagram. The file index (IVF) is created in one pass. For each chunk C$_i$ (here 4 tokens) we get one neighbor. To generate the token for chunk $C_2$, we utilize the neighbor of chunk $C_1$. For \name layers 1-5,7,8,10,11 there is no chunked cross-attention (CCA) block. The information flows directly from the GPT-2 attention block to the FFW layer.}
\vspace{0.5cm}
\label{fig:arch-retro-diagram}
\end{figure*}

\textbf{1) We enable RAG to work with a small-scale database.} In \name, we change the embedding of the retrieval system, and for the first time add regularization in the form of Gaussian noise to the neighbor embeddings of the non-parametric memory. We employ a novel embedding model that excels in semantic similarity search allowing retrieval of higher quality neighbors despite searching in a \emph{sparse} non-parametric memory, and the regularization consistently improves domain shift generalization. We show that the new retrieval helps with language modeling on small-sized retrieval databases over having no retrieval database.

\textbf{2) We advocate for the use of in-memory computing (IMC) hardware to reduce the complexity of search.} Analog IMC performs certain operations in place in-memory by exploiting the physics of e.g., non-volatile memory devices, offering $O(1)$ computational complexity for similarity search with a set of precomputed vectors (see~\citep{IMC_NatureNano2020} for an overview). This feature is highly valuable for the efficient realization of memory-augmented neural networks as shown in~\citep{Karunaratne2021}. Due to nonidealities, however, the analog search operation of IMC is noisy. We simulate this behavior of the IMC hardware by adding a wide range of noise to the neighbor embeddings at inference time. We show that the resulting noisy retrieval does not decrease the \name's performance (i.e., a maximum of <1\% drop) thanks to our training with regularization. This suggests deployment of the non-parametric memory on the IMC hardware which would significantly improve inference time, in addition to energy consumption and computational density.

\textbf{3) For domain shift generalization, we plug-and-play the related domain database without any fine-tuning in \name.} This is encouraged by having a closer look at the \textsc{Retro} that reveals that size is not the most important aspect of the retrieval database: \emph{a smaller but relevant retrieval database improved the model performance w.r.t. an order of magnitude larger database.} Although \textsc{Retro} advocates for a large retrieval database with as much variety as possible to make the model more general, we show that their system can be even more modular than that. In \name, it is straightforward to replace the retrieval database with data from domain A if inference sequences are also from domain A. With this high degree of plug-and-play, there is no need to build a database with trillions of tokens from as many sources as possible. The user can easily build a new database for each domain of interest. Furthermore, we show \name can also benefit from fine-tuning similar to the other models without a retrieval.


\section{\name}
\label{sec:implementing_retro}
Our proposed \name is a medium-sized parametric model based on \textsc{Retro} with a small-scale non-parametric database. In \name, the embedding model used for the neighbor search is a \textsc{Bert} based sentence similarity model called \textsc{SBert}. The architecture diagram of \name is shown in Figure~\ref{fig:arch-retro-diagram}, where we enhanced GPT-2 attention blocks similarly to what has been down with \textsc{Retro}-fitting to improve our language modeling performance without extensive re-training. \textsc{Retro}-fitting as introduced in the \textsc{Retro} paper~\citep{retro}, refers to taking a language model not trained with retrieval and adding retrieval in the form of a frozen retrieval database and chunked cross-attention (CCA) blocks in order to incorporate the information from neighbor embeddings in the training sequences.

To query the key-value retrieval database (DB), each input sequence is split into chunks. An input sequence contains 1024 tokens, for a chunk length of 64 tokens this is 16 chunks per sequence. 
\begin{equation}\label{eq:chunking}
DB([Chunk_0,...,Chunk_{15}]) = 
\begin{bmatrix}
[N_0,C_0]_0, ... ,[N_{9},C_{9}]_{0}\\
 \vdots \\
[N_0,C_0]_{15}, ... ,[N_{9},C_{9}]_{15}\ 
\end{bmatrix}
\end{equation}
The neighbors consist of [N, C], each N containing 64 tokens and C containing another 64 tokens. The retrieval is based on N only, the key of this key-value pair, the value being C, which is the continuation of N. We retrieve them for each chunk. Equation~\ref{eq:chunking} shows retrieval of 10 neighbors for each chunk in a sequence consisting of 16 chunks from the retrieval database (DB).

It is worth noting that \textsc{Retro} does not exhibit any meaningful results concerning improved performance below a retrieval database size of 100 billion tokens. For their two smallest models, adding a retrieval database of two billion tokens improves performance slightly (0.1 bits-per-byte on C4). However, then even increasing the database ten-fold does not change performance further. The only meaningful jump in performance for the smallest models happens once the retrieval database increases from 360\,B tokens to 900\,B tokens. This is expensive in terms of space and compute power, thus unrealistic for most setups. Instead, in \name, we show how performance can be improved using retrieval databases at orders of magnitude smaller scale (570\,K up to 2.89\,B database tokens) through architectural enhancements and training strategies. 

For \name we use \textsc{Faiss}~\citep{faiss}, an inverted vector files (IVF) index, while \textsc{Retro}'s retrieval database is built using ScaNN~\citep{scann}, which also builds a vector index, but uses asymmetric hashing to obtain the candidate list for a query vector. \textsc{Retro} embed their chunks using \textsc{Bert} embeddings~\citep{bert}, while for \name we choose \textsc{SBert}~\citep{sentencetransformer}, specifically trained for semantic similarity search. For the tokenization, \textsc{Retro} uses SentencePiece~\citep{sentencepiece}, a byte-level encoding model, and a vocabulary size of 128'000. \name instead uses the GPT-2 tokenizer, thus a vocabulary size 50'257.

\subsection{Embeddings for Neighbor Search}
\label{sec:word_embeddings}
In \name, we set out to work with small retrieval databases and almost no token overlap (see Appendix \ref{app:full_analysis_on_overlap}). Thus, we have to be judicious about which neighbors are returned. In this setup with smaller retrieval databases, the search space of sequences is not as densely populated, thus not finding the true closest sequence carries a bigger penalty. 
Considering both ease of use and theoretical results, we choose to use \textsc{SBert}~\citep{sentencetransformer}, a \textsc{Bert}-based sentence-similarity model, in \name.

We use the pre-trained model \textit{multi-qa-mpnet-base-dot-v1} of \textsc{SBert} due to its superior performance in semantic textual similarity tasks in the massive text embedding benchmark (MTEB~\citep{muennighoff2023mteb}). The embedding model outputs 768-dimensional vectors and has a model size of 420~MB. This vector dimensionality and model size are compatible with the embedding size and the model size of the GPT-2 backbone used in \name. 

\textsc{SBert} adds a pooling operation to the output of Bert, resulting in a fixed-size sentence embedding. 
For our chosen embedding model, this pooling operation is \textit{CLS pooling}, which involves using the CLS token embedding to represent the sentence. These embeddings were specifically trained for semantic similarity search. More details on \textsc{SBert} can be found in Appendix~\ref{app:background_and_related_work}.

\subsection{Regularization of Non-Parametric Memory}
\label{sec:regularization}
The role of noise in the retrieval has to the best of our knowledge never been analyzed. Here, a first foray into this topic is made. Adding noise to word embedding as a way of improving the generalization capabilities of language models has been done before~\citep{zhu2020freelb}, but not in combination with retrieval.

Taking inspiration from NEFTune~\citep{neftune}, we aim to improve generalization through a word-embedding regularizer. We introduce a new regularization method by adding noise to the word embeddings of the non-parametric memory, in contrast to NEFTune which regularizes the input sequence, in \name their retrieved neighbors are regularized (see Figure~\ref{fig:arch-retro-diagram}). 

For a matrix M of dimension n\_emb $\times$ seq\_len describing the embedding matrix of a sequence, we compute:
\begin{equation}\label{eq:imc_noise}
\sigma = \lambda_t \cdot MEAN(ABS(M))
\end{equation}
The relative standard deviation $\lambda_t$ describes the magnitude of the noise added. During training, we add a vector sampled from this $\mathcal{N}(0,\sigma)$ distribution to our neighbor embeddings. 

We explore a variety of regularizers, some similar to NEFTune, some more similar to the intrinsic noise in the stochastic IMC hardware (see Section~\ref{sec:result_regularization}), ensuring that the signal-to-noise ratio from different approaches remains comparable.

\section{Results}
\label{sec:results}
\subsection{Experimental Setup}
We base our work on an earlier \textsc{Retro} implementation~\citep{labml} in which the retrieval database and the similarity search are implemented with \textsc{Faiss} index. The training is accelerated with a single Tesla V100 GPU with a maximum 80~GB memory.

\subsection{Language Modeling}
\label{sec:setup}

The task is language modeling on WikiText-103 and the numbers we report are the perplexities for WikiText-103-Validation. Just as in \textsc{Retro}, we employ a sliding window approach, where we compute the perplexities for an overlapping proportion greater than or equal to 75\% of the context, more on this in Appendix~\ref{app:perplexity_discussion}.

Our work is also inspired by the Chinchilla law~\citep{chinchilla} to determine the architecture and training strategy so that it fits a setup of any scale.
For this particular task, we train on one GPU for 1’078’012 sample sequences resulting in a total of $1.104 \times 10^9$ tokens. The time to train one epoch grows with the number of neighbors, but sub-linearly. Training on 10 neighbors takes about twice as long as training on 2 neighbors but training on 2 neighbors takes about the same time as training on no neighbors.

We conduct an initial exploration of the experiment space with only a frozen GPT-2 component. So we train not only the CCA but also the feed-forward (FFW) blocks (see Appendix~\ref{app:architecture} for the corresponding architecture diagram). A pattern emerges, where \textsc{SBert} does better than \textsc{Bert} and 3 neighbors do best while not taking much more time to train on, compared to no neighbors. 

\subsubsection{Retrieval}
\label{sec:retrieval}
For our retrieval experiments we only freeze the GPT-2 layers as shown in Figure \ref{fig:arch-retro-diagram}. We explored un-freezing these layers as well, see Appendix~\ref{app:additional_ablations}. Both \name-off and \name-on are trained on the same data and, aside from the GPT-2 attention blocks and embeddings, from scratch.

\subsubsection{Ablation of Embeddings for Neighbor Search}
For the first batch of experiments, we gradually increase the number of neighbors from 2 up to 10 and run the experiments for three random seeds. In Table~\ref{tab:embedding_ablation} we report their average. \textsc{Bert} embeddings with 5 neighbors already outperforms \name-off. Moreover, changing the embedding model to \textsc{SBert} improves our performance significantly. This shows that not only does \textsc{SBert} find more semantically similar neighbors, but our model also correctly attends to them. 

Furthermore, we see that for \textsc{Bert} embeddings all cases except the number of neighbors equal to 5 under-perform \name-off. At this stage of the model training, where we only freeze GPT-2 blocks, it is too difficult for our model to find its way through the loss landscape to get to the best possible performance. Adding sub-optimal neighbors at this point exacerbates the issue. 

This makes the \textsc{SBert} results even more remarkable. The neighbors found through the \textsc{SBert} embeddings genuinely inform the model, helping it to converge.
Still, not all numbers of neighbors improve the performance compared to no retrieval, confirming what~\citep{selfrag} and~\citep{toolformer} have already observed. 

Finally, we note that with this setup, for the \textsc{Bert} embeddings, and \name-off the variance between the random seeds is very large. This is partially because we were unable to train all checkpoints to convergence, as some got stuck in local minima. Thus another aspect we show here is how \textsc{SBert} embeddings help to avoid/escape these local minima.

\begin{table}[t]
\vspace{0.2cm}
\caption{Perplexity results of experiments for \name-on/-off with \textsc{Bert} and \textsc{SBert} embeddings averaged over three random seeds.}
\vspace{0.2cm}
\label{tab:embedding_ablation}
\small
\centering
\ra{1.3}
\begin{tabular}{lllll}
\toprule
{\# Neighbors} & {\textsc{Bert}} & {\textsc{SBert}}\\
\midrule
{2}        & 50.59           & 24.05\\
{3}        & 44.93           & \textbf{22.61}\\
{5}        & \textbf{28.92}  & 35.11\\
{7}        & 49.11           & 34.08\\
{10}       & 49.10           & 23.24\\
{\name-Off}& 32.06           & 32.06\\
\bottomrule
\end{tabular}
\end{table}

Overall, adding neighbors at a stage where the language model itself has not converged yet can help, but does not always. We conclude that we must first address the language modeling before we can add neighbors and benefit from them.

\subsubsection{Ablation of Regularizations}
\label{sec:result_regularization}

\begin{table*}[ht]
\caption{Perplexity of the plug-and-play domain shift experiments with six random seeds for different regularizers: $\lambda_t$ refers to the relative standard deviation of a Gaussian regularizer, $\alpha$ refers to the NEFTune uniform regularizer, N/A refers to no regularizer. $\lambda_i$ refers to the relative standard deviation of the additive noise that simulates different IMC hardware inducing noisy retrieval at inference time.}
\vspace{0.25cm}
\centering
\ra{1.3}
\small
\begin{tabular}{lllllllll}
\toprule
$\lambda_i$ & Regularizer       &
    {BBC-News} &
    {Reuters} &
    {\begin{tabular}[c]{@{}l@{}}Founding \\ docs\end{tabular}}&
    {\begin{tabular}[c]{@{}l@{}}CNN-\\ DailyMail\end{tabular}}&
    {\begin{tabular}[c]{@{}l@{}}Atticus \\ Contracts\end{tabular}} &
    {\begin{tabular}[c]{@{}l@{}}Open-\\ WebText\end{tabular}} &
    {\begin{tabular}[c]{@{}l@{}}Slim-\\ Pajama\end{tabular}} \\
    \midrule
\multirow{4}{*}{0}        & N/A            & $127.82^{\pm 1.9}$  & $154.73^{\pm2.2}$ & $260.63^{\pm4.2}$                                       & $167.21^{\pm1.4}$                                        & $316.48^{\pm4.9}$                                           & $299.81^{\pm3.2}$                                        & $475.23^{\pm4.2}$                                      \\
         & $\alpha = 10$  & $127.84^{\pm2.3}$  & $154.73^{\pm2.0}$ & $260.23^{\pm7.4}$                                        & $167.26^{\pm2.6}$                                        & $318.01^{\pm5.8}$                                        &   $300.87^{\pm7.6}$                                        & $477.7^{\pm7.4}$                                       \\
         & $\lambda_t = 0.2$  &\boldmath{$127.73^{\pm0.8}$}  & \boldmath{$153.16^{\pm2.0}$ }& \boldmath{$258.27^{\pm3.4}$}                                       & \boldmath{$166.25^{\pm2.7}$ }                                       & \boldmath{$314.28^{\pm7.7}$ }                                          & \boldmath{$297.08^{\pm5.6}$}                                       & \boldmath{$472.87^{\pm8.3}$}                                      \\
         & $\lambda_t = 0.4$ & $128.10^{\pm1.8}$  & $154.57^{\pm3.0}$ & $261.7^{\pm6.2}$                                        & $167.32^{\pm2.5}$                                        & $317.94^{\pm6.0}$                                           & $302.20^{\pm8.0}$                                          & $477.42^{\pm10.6}$                                     \\
\midrule
\multirow{4}{*}{0.2}      & N/A            & $127.86^{\pm1.9}$  & $154.75^{\pm2.2}$ & $260.68^{\pm4.1}$                                        & $167.24^{\pm1.4}$                                        & $316.54^{\pm4.8}$                                           & $299.88^{\pm3.1}$                                       & $475.36^{\pm4.0}$                                      \\
         & $\alpha = 10$      & $127.88^{\pm2.2}$  & $154.76^{\pm1.9}$ & $260.26^{\pm7.4}$                                        & $167.3^{\pm2.5}$                                         & $318.17^{\pm5.8}$                                           & $300.99^{\pm7.6}$                                        & $477.79^{\pm7.4}$                                      \\
         & $\lambda_t = 0.2$  & \boldmath{$127.75^{\pm0.8}$}  & \boldmath{$153.18^{\pm2.0}$} & \boldmath{$258.34^{\pm3.4}$}                                       & \boldmath{$166.27^{\pm2.6}$}                                        & \boldmath{$314.35^{\pm7.7}$}                                           & \boldmath{$297.18^{\pm5.6}$}                                       & \boldmath{$473.02^{\pm8.3}$ }                                     \\
         & $\lambda_t = 0.4$ & $128.13^{\pm1.8}$  & $154.62^{\pm3.0}$ & $261.8^{\pm6.3}$                                        & $167.35^{\pm2.5}$                                         & $318.06^{\pm6.0}$                                           & $302.33^{\pm8.1}$                                       & $477.71^{\pm10.7}$                                     \\
\midrule
\multirow{4}{*}{0.4}      & N/A            & $127.94^{\pm2.0}$  & $154.84^{\pm2.1}$ & $260.86^{\pm3.9}$                                       & $167.35^{\pm1.3}$                                        & $316.83^{\pm4.6}$                                           & $300.1^{\pm2.9}$                                        & $475.76^{\pm3.4}$                                      \\
         & $\alpha = 10$ & $127.98^{\pm2.2}$  & $154.87^{\pm1.9}$ & $260.41^{\pm7.4}$                                       & $167.37^{\pm2.5}$                                        & $318.39^{\pm5.8}$                                           & $301.24^{\pm7.6}$                                       & $478.18^{\pm7.4}$                                      \\
         & $\lambda_t = 0.2$  &\boldmath{$127.83^{\pm0.8}$}  & \boldmath{$153.31^{\pm1.9}$ }& \boldmath{$258.58^{\pm3.4}$ }                                      & \boldmath{$166.35^{\pm2.5}$}                                        & \boldmath{$314.6^{\pm7.7}$ }                                           & \boldmath{$297.47^{\pm5.4}$}                                       & \boldmath{$473.45^{\pm8.2}$}                                      \\
         & $\lambda_t = 0.4$ & $128.30^{\pm1.9}$  & $154.81^{\pm3.1}$ & $262.11^{\pm6.4}$                                       & $167.46^{\pm2.6}$                                        & $318.44^{\pm6.0}$                                           & $302.76^{\pm8.3}$                                       & $478.73^{\pm11.0}$                                     \\
\midrule
\multirow{4}{*}{1.0}      & N/A            & $128.49^{\pm2.2}$  & $155.54^{\pm3.3}$ & $261.92^{\pm3.4}$                                        & $168.53^{\pm0.9}$                                        & $318.66^{\pm3.7}$                                            & $301.63^{\pm2.1}$                                       & $478.18^{\pm1.5}$                                      \\
         & $\alpha = 10$      & $128.85^{\pm2.2}$  & $155.94^{\pm3.5}$ & $262.07^{\pm7.3}$                                       & $168.17^{\pm2.3}$                                        & $321.32^{\pm5.4}$                                           & $303.92^{\pm7.4}$                                       & $482.51^{\pm6.3}$                                      \\
         & $\lambda_t = 0.2$  & \boldmath{$128.36^{\pm0.8}$} & \boldmath{$154.06^{\pm2.4}$ }& \boldmath{$259.88^{\pm3.8}$}                                       & \boldmath{$166.87^{\pm2.1}$ }                                       & \boldmath{$316^{\pm8.5}$}                                               & \boldmath{$299.16^{\pm5.5}$}                                       & \boldmath{$475.97^{\pm8.5}$}                                      \\
         & $\lambda_t = 0.4$ & $129.42^{\pm0.7}$ & $156.42^{\pm4.6}$ & $265.03^{\pm7.1}$                                       & $168.69^{\pm3.7}$                                        & $322.21^{\pm6.0}$                                           & $306.4^{\pm9.6}$                                        & $504.2^{\pm14.6}$                                     \\
         \bottomrule
         \\
\end{tabular}
\label{tab:ppl_per_inferencenoise}
\end{table*}

Similarly to NEFTune, we add uniform noise to the word embeddings during training. The goal of it is to improve the generalization capabilities of our model. We add it to the training sequence itself, to the neighbors, and also both. We again average the result over three random seeds and ablate the noise type as well as the noise strength.

Preliminary ablations performed to determine the best type of noise, both in terms of magnitude and in terms of where the noise is added (so whether to add on the input embeddings or to the neighbor embeddings), showed that regularization works best when moderately added only to the neighbor embeddings. The full results are in Appendix~\ref{app:regularization}. We also performed additional experiments to determine the best type of noise among the uniform and the Gaussian while ensuring a similar signal-to-noise ratio. Results are shown in Table~\ref{tab:ppl_per_inferencenoise}. A Gaussian with zero mean and $\lambda_t = 0.2$ (the variance as described by Equation~\ref{eq:imc_noise}) improves our generalization performance better than the rest of the regularizers we have tried. 

Table~\ref{tab:domain_shift_experiments} shows the impact of this best-performing regularizer (the Gaussian regularizer with $\lambda_t = 0.2$) in both language modeling and domain shift.
In the WikiText-103 language modeling, the regularizer improves or at least does not impair the validation perplexity in various inference settings: having no neighbors, ideal retrieval, or different amounts of noise during inference (i.e., $\lambda_i \in \{0.2,0.4,1\}$). 
For the domain shift experiments, it exhibits more promising behavior and consistently shows benefits when different noisy scenarios are considered.
More details are provided in the next subsection.
%

\subsection{Domain Shift Generalization: Plug-and-play}
\label{sec:domainshift_plug_and_play}
\begin{table*}[t]
\caption{Perplexity results for the language modeling and the plug-and-play domain shift experiments for different model inference settings, averaged over six random seeds. Models with various inference settings are evaluated: \name-off and \name-on (without neighbors, with ideal neighbors, and with noisy neighbors impacted by $\lambda_i \in \{0.2,0.4,1\}$). Models are trained without a regularizer (None) and with our best regularizer (Gaussian with $\lambda_t = 0.2$).}
\vspace{0.25cm}
\label{tab:domain_shift_experiments}
\centering
\ra{1.3}
\small
\begin{tabular}{@{}lllcccccccc@{}}
\toprule
 &
   &
   &
   
  \rotatebox[x=0pt,y=-14pt]{90}{WikiText} &
  \rotatebox[x=0pt,y=-14pt]{90}{BBC-News} &
  \rotatebox[x=0pt,y=-14pt]{90}{Reuters} &
  \rotatebox[x=0pt,y=-14pt]{90}{\begin{tabular}[c]{@{}l@{}}Founding \\ docs\end{tabular}}&
  \rotatebox[x=0pt,y=-14pt]{90}{\begin{tabular}[c]{@{}l@{}}CNN-\\ DailyMail\end{tabular}}&
  \rotatebox[x=0pt,y=-14pt]{90}{\begin{tabular}[c]{@{}l@{}}Atticus \\ Contracts\end{tabular}} &
  \rotatebox[x=0pt,y=-14pt]{90}{\begin{tabular}[c]{@{}l@{}}Open-\\ WebText\end{tabular}} &
  \rotatebox[x=0pt,y=-14pt]{90}{\begin{tabular}[c]{@{}l@{}}Slim-\\ Pajama\end{tabular}} \\ \cmidrule(l){4-11} Retrieval 
 & Settings &
  Regularizer & & & & & & & & \\ \midrule
Off & Normal inf. & None & 20.62 & 130.31 & 168.91 & 266.85 & 171.56 & 329.98 & 312.15 & 489.46 \\ \midrule
\multirow{10}{*}{On} &
  \multirow{2}{*}{\begin{tabular}[c]{@{}l@{}}No neighbors \end{tabular}} &
  None &  19.74 &  129.63 &  157.30 &  265.53 &  170.43 & 322.67 &  306.10 &  485.84 \\
 &   
 &  
 Gaussian &  \textbf{19.72} & \textbf{129.17}  & \textbf{155.05} & \textbf{260.16} & \textbf{168.78} & \textbf{318.37} & \textbf{301.94} &  \textbf{481.10}\\ \cmidrule(l){2-11} 
 &
  \multirow{2}{*}{\begin{tabular}[c]{@{}l@{}}Ideal retrieval, \\ $\lambda_i=0$ \end{tabular}
  } &
  None &  19.60 &  127.82 &  154.73 &  260.63 &  167.21 &  316.48 &  299.81 & 475.23 \\
 &
 &
  Gaussian & \textbf{19.60} & \textbf{127.73} & \textbf{153.16} & \textbf{258.27} & \textbf{166.25} & \textbf{314.28} & \textbf{297.08} & \textbf{472.87} \\ \cmidrule(l){2-11} 
 &
  \multirow{2}{*}{\begin{tabular}[c]{@{}l@{}}Noisy retrieval, \\ $\lambda_i=0.2$ \end{tabular}} &  
  None &  19.60 &  127.86 &  154.75 &  260.68 &  167.24 &  316.54 &  299.88 &  475.36 \\
 &
 &
  Gaussian & \textbf{19.60} & \textbf{127.75} & \textbf{153.18} & \textbf{258.34} & \textbf{166.27} & \textbf{314.35} & \textbf{297.18} & \textbf{473.02}\\ \cmidrule(l){2-11} 
 &
  \multirow{2}{*}{\begin{tabular}[c]{@{}l@{}}Noisy retrieval, \\ $\lambda_i=0.4$\end{tabular}} &
  None &  19.61 &  127.94 &  154.84 &  260.86 &  167.35 &  316.83 &  300.10 &  475.76 \\
 
 &
 &
  Gaussian & \textbf{19.60} & \textbf{127.83} & \textbf{153.31} & \textbf{258.58} & \textbf{166.35} & \textbf{314.60} & \textbf{297.47} & \textbf{473.45} \\ \cmidrule(l){2-11} 
 &
  \multirow{2}{*}{\begin{tabular}[c]{@{}l@{}}Noisy retrieval, \\ $\lambda_i=1.0$\end{tabular}} &
  None &  19.66 &  128.49 &  155.54 &  261.92 &  168.53 &  318.66 &  301.63 &  478.18 \\
 &
 &
  Gaussian & \textbf{19.63} & \textbf{128.36} & \textbf{154.06} & \textbf{259.88} & \textbf{166.87} & \textbf{316.00} & \textbf{299.16} & \textbf{475.97}\\
  \bottomrule
  \\
\end{tabular}
\end{table*}

We use three architectural settings for domain shift generalization:
\begin{itemize}
    \item Off (i.e., no retrieval). This model was never trained with retrieval.
    \item \name-on with no neighbors: This model was trained with retrieval, but we do not give it any neighbors at inference time.
    \item \name-on with ideal (i.e., not noisy) retrieval: This model was trained with retrieval and we give it the ten nearest neighbors per chunk at inference time.
\end{itemize}
The setting \name-on with \textit{no neighbors} is meant to ascertain how much of the domain shift performance gain is simply due to improved language modeling capabilities, and not due to the retrieval itself. In settings with \name-on, we further have a model trained with a regularizer or without one.
In Table~\ref{tab:domain_shift_experiments}, we only include the best-performing regularizer, a Gaussian regularizer with $\lambda_t = 0.2$ (see Equation~\ref{eq:imc_noise}), which also reflects the signal-to-noise ratio observed in the modern IMC hardware (more on this in Section~\ref{sec:hardware-aware_regularization}).

For domain shift experiments with \name-on settings, we take the model and plug in a new retrieval database. In this case, creating a new retrieval database consists of the following steps. First, we train the models on a base dataset A. For a new domain dataset B, C, D, or E (i.e., a different domain) we create a retrieval database with the training tokens of B, C, D, or E, respectively, then chunk the validation sequences of B, C, D, or E, respectively, and find the closest neighbors in the retrieval database for each validation chunk. Then, we plug this database, which includes chunk-to-neighbor mapping, into our model, feed it with the validation sequences as input, and measure the new perplexity. We do not fine-tune the models on any of the target domains whatsoever. 

\begin{table}[ht]
\vspace{0.2cm}
\caption{Size of validation data and retrieval database in thousands of tokens and database entries in thousands for datasets considered for domain shift experiments.}
\vspace{0.2cm}
\label{tab:domainshift_data}
\small
\centering
\ra{1.3}
\begin{tabular}{lwr{0.8cm}wr{1.5cm}wr{1.5cm}}
\toprule
{Dataset} &
  {\begin{tabular}[c]{@{}l@{}r@{}}\# val \\{tokens}\\'000\end{tabular}} &
  {\begin{tabular}[c]{@{}l@{}r@{}}\# {tokens} in\\ retrieval DB\\'000\end{tabular}} &
  {\begin{tabular}[c]{@{}l@{}r@{}}\# entries in \\ retrieval DB\\'000\end{tabular}} \tabularnewline  \midrule
{BBC-News}                         & 461    & 570       & 8.9      \tabularnewline 
{Reuters}              & 174    & 3’455     & 54     \tabularnewline 
{Founding docs} & 18’606 & 55’819    & 872    \tabularnewline 
{CNN-DailyMail}                    & 10’116 & 223’216   & 3’488  \tabularnewline 
{Atticus contracts}  & 7’606 & 456’682  & 7’136  \tabularnewline 
{OpenWebText}                      & 13’914 & 2’477’892 & 38’717 \tabularnewline 
{SlimPajama}                       & 4’948  & 2’886’850 & 45’107 \\
\bottomrule
\end{tabular}%
\end{table}

Table~\ref{tab:domainshift_data} lists the datasets used for domain shift experiments, which are ordered by the size of the retrieval databases. The datasets are chosen such that they are sourced from a variety of domains and have sufficiently distinctive characteristics to identify them as unique domains. The datasets are further explained with their differences highlighted in Appendix~\ref{appendix:datasets}.

In the context of this work, a domain refers to the formality of the language. At one end, we have Wikipedia, which is highly formal text, and on the other end, we have SlimPajama, which consists mostly of text gathered through web crawling and thus contains predominantly informal, poorly or unstructured text, in a variety of languages and even quite a bit of code (see Appendix~\ref{app:slimpajama} for more details). 

Unlike for the previous experiments, here we take our best baseline transformer (\name-off), re-set the optimizer and only train the CCA blocks on top of it using dataset A (see above). For the neighbors, we choose three neighbors and \textsc{SBert} embeddings. This way, we can more closely emulate the experiments done by \textsc{Retro} and observe what the true performance gain of retrieval is. This has the benefit of not only decreasing the variance across seeds but additionally enabling us to keep the \name-off performance exactly intact.  

We take the best \name-off checkpoint and keep training it for six random seeds, with and without retrieval, and with and without regularization. WikiText-103 is taken as the base dataset (dataset A, see above) for the taining. We train both \name-on and \name-off, in order to confirm that the improved performance is not due to the increased number of training tokens. As is evident in the rows for \name-on with ideal retrieval in Table~\ref{tab:domain_shift_experiments}, when compared to Table~\ref{tab:embedding_ablation} with training from scratch, this setup decreases WikiText-103-Validation set perplexity for both, \name-on and \name-off. The seemingly significant perplexity improvement of \name-off is due to there no longer being one particularly bad random seed, as this setup reduces the variance across seeds. 

\name-off has 109\,M trainable parameters, and 234\,M parameters in total, while \name-on has 24\,M additional parameters due to the CCA blocks. Moreover, since \name-on freezes all parameters except for the CCA blocks, only those 24\,M are updated, as opposed to 109\,M parameters for \name-off. 

As shown in Table~\ref{tab:domain_shift_experiments}, even without a regularizer, we see across seeds and datasets that \name-on deals with the domain shift better than \name-off. Adding a regularizer to the retrieval improves this performance even further. In this table, the datasets are ordered by the size of the retrieval databases. Although retrieval and regularization help, the perplexity also keeps going up, despite our retrieval database size increasing as well. However, considering the improved percentage, where \name-off is our baseline, we can see that our \name-off model struggles with a large domain shift, where retrieval and regularization help the most. 

This performance improvement is not solely due to the fact that \name-on is already better at language modeling. Consider the "No neighbors" rows, where we replace the nearest neighbors of a sequence with the sequence itself and mask out the continuation. Hence, the model cannot benefit from retrieval in this setup (i.e., no extra information is retrieved) but includes the additional 24\,M parameters due to the CCA blocks. We observe that \name-on's performance without neighbors drops significantly and the performance is similar to \name-off. This indicates that neither the additional parameters nor the better language modeling capability, but the actual retrieval improves our model's generalization ability. 

For \name-on with the Gaussian regularization we observe that the model outperforms \name-on without a regularizer on all domain shift datasets. This suggests that the retrieval model can be trained with the regularizer to emulate a denser search space. Moreover, in the setup where no neighbors are given, the regularizer also improves the performance of the language model without any access to retrieval.

\subsubsection{Noisy retrieval with IMC hardware}
\label{sec:hardware-aware_regularization}
\begin{table}[b]
\centering
\caption{Average search time (ms) per dataset on a Tesla V100 GPU using \textsc{Faiss} index.}
\label{fig:faiss_table}
\ra{1.3}
\begin{tabular}{lllll}
\toprule
\multirow{2}{*}{Dataset} & \multirow{2}{*}{\begin{tabular}[c]{@{}l@{}}\# db entries\\ in millions\end{tabular}} & \multicolumn{3}{c}{Search time {[}ms{]}} \\ \cline{3-5} 
              &       & Average & Max     & Min    \\
\midrule
WikiText-103  & 1.88  & 76.97   & 93.96   & 56.31  \\
CNN-DailyMail & 3.49  & 132.49  & 170.46  & 93.38  \\
SlimPajama    & 45.11 & 896.22  & 1237.13 & 724.63     \\
\bottomrule
\end{tabular}
\end{table}
Before delving into the potential benefits and challenges of retrieval with the IMC hardware, let us explain how it has been done in the state-of-the-art. Currently, we search for the nearest neighbors using \textsc{Faiss} which is an inverted vector file (IVF) index. An IVF index clusters the vectors to be searched into $c$ centroids, where each vector is assigned to one centroid. Upon receiving a query, \textsc{Faiss} searches $nprobe$ of those $c$ cells to find the nearest neighbors. This reduces the search time by reducing the search space to the number of centroids to be probed while preserving good performance. However, an IVF index requires training to identify optimal centroids.

In Table~\ref{fig:faiss_table}, we measure the time it takes for the search function call to \texttt{IndexIVF\_search} to complete. This function is called on 16 chunks concurrently on a Tesla V100 GPU. More on these measurements including the search time distributions can be found in Appendix~\ref{app:faiss}.

Next, as a natural extension of memory-augmented neural networks~\citep{Karunaratne2021}, we assess whether the non-parametric memory of \name can be moved to a specialized hardware that operates based on the IMC principles. This would make vector searches and thus retrieval much faster, as it foregoes the memory wall issues prevailing in GPUs caused by the high bandwidth data transfers. If the retrieval database is very large (such as in \textsc{Retro}'s case where we have trillions of tokens and 28 billion database keys) the data transfer becomes the bottleneck, especially during inference. We estimate that on a larger scale IMC hardware platform inspired by the early-stage prototype~\citep{khaddam2021hermes}, the similarity search time could be brought down to several hundred nano-seconds.

One drawback of the IMC hardware however is low-precision and noisy similarity searches due to analog computations with non-idealities. We simulate the noise on such a hardware platform by a Gaussian distribution with zero mean and a certain standard deviation $\sigma$, described by Equation~\ref{eq:imc_noise}. This additive noise at inference time is denoted by $\lambda_i$ as opposed to $\lambda_t$ which is used to describe the noise during training. For a more in-depth explanation of noise modeling on IMC-based hardware, please refer to~\citep{analogai-hwkit}. For instance, for a recent large-scale chip based on phase-change memory devices~\citep{khaddam2021hermes}, this relative standard deviation is 0.2. 

Adding Gaussian noise with a variety of relative standard deviations to our neighbor word embeddings at inference time gives us stable results, despite it not being seen during training (i.e., trained with $\lambda_t=0.2$ while tested with $\lambda_i \in \{0,0.2,0.4,1.0\}$ ). This is evident in Table~\ref{tab:domain_shift_experiments}, in the rows describing the setting \textit{noisy retrieval} with no regularize (i.e., \textit{None}). Even for a relative standard deviation as large as 1.0, the performance never drops more than 1\% compared to the ideal retrieval without any noisy neighbors. This suggests that the searches on the non-parametric memory can be moved to the IMC hardware without issue. 

Training with the regularizer does not decrease the performance drop upon adding noisy retrieval at inference time in all cases. For  $\lambda_i=0.2$ and  $\lambda_i=0.4$, even when we did not train using a regularizer, our performance dropped about the same relative to ideal retrieval as if we had trained with a regularizer. On the other hand, for $\lambda_i=1.0$, a large relative standard deviation, this regularized training does make a difference. Where other models fail, the one trained with a Gaussian regularizer is barely affected. This can be seen more clearly in Table~\ref{tab:ppl_per_inferencenoise} where we also add the standard deviation error bar for the six random seeds. 

Finally, we perform an ablation study with a more detailed and accurate noise inference model given in Analog AI hardware kit~\citep{analogai-hwkit}. This model splits the database to store its content on realistic IMC crossbar sizes. During the similarity search, it captures the non-idealities associated with the crossbars both short term such as read noise, and long term such as drift. For the memory device level noise modeling Phase-change Memory (PCM) preset~\citep{pcm_noise_model} is chosen with a maximum conductance of 25~uS and the rest of the parameters as default.

\begin{table}[!ht]
\centering
\caption{Inference perplexity with and without the accurate noisy retrieval}
\label{tab:aihwkit_nosiyret}
\ra{1.3}
\begin{tabular}{l|l|l}
\hline
Dataset                   & Retrieval                 & Perplexity \\ \hline
\multirow{2}{*}{WikiText} & ideal                     & 19.7872    \\
                          & \textbackslash{}w aihwkit~\citep{analogai-hwkit} & 19.7876    \\ \hline
\multirow{2}{*}{BBC-News} & ideal                     & 62.9104    \\
                          & \textbackslash{}w aihwkit~\citep{analogai-hwkit} & 62.9118 \\ 
\hline
\end{tabular}
\end{table}
The results of this study are presented in Table~\ref{tab:aihwkit_nosiyret}. We use the best-trained models on the WikiText and BBC-News datasets and compare the inference results where in one case there is no noise on retrieved neighbors and in the other case the retrieval neighbors' noise is modeled comprehensively using the Analog AI hardware kit. We notice that the noisy retrieval across both datasets drops only by a negligible margin (<0.003\%) indicating the robustness of the \name models trained with regularization towards noisy retrieval.

\begin{table*}[!ht]
\caption{\name-off generated text for a SlimPajama-6B-Validation sample with the lowest \name-off perplexity.}
\label{tab:continuation-off-slimpajama-onlybestgeneration}

\begin{tabularx}{1\textwidth}{p{0.43\textwidth}p{0.235\textwidth}p{0.255\textwidth}}
\hline
Context                                                                                                                                                                                                                                                                                                                                                                                                                                                                                                                                                  & Real Continual Chunk {[}1.0{]}                                                                                                                                                                                                                                                        & Top-p, p=0.9 {[}0.403{]}                                                                                                                                                                                                                                                                                  \\ \hline
\begin{scriptsize}
(...) he said besides this the government could give freight subsidy and storage subsidy on the pattern of west bengal and uttar pradesh governments. mr majithia said seed potato could also be exported to pakistan, afghanistan and sri lanka. "if all these steps are are not taken the government alone will be responsible for driving farmers to suicide in despair ", he added.asserting that the state was staring at a farm crisis, the akali leader said potato prices had hit a rock bottom of rs 3 per kilogram. he said in such a situation farmers \end{scriptsize}
& 
\begin{scriptsize} were unable to recover the cost of production which came to between rs 6 to rs 8 per kg after adding harvesting and packing costs. "many farmers are even putting off harvesting of their crop due to this reason", he added.the shiromani akali dal (sad) today demanded a relief package of 
\end{scriptsize}
& 
\begin{scriptsize} pubabout. he said before,... then the state government is able to attract customers, prevent famine, preventing corruption and corruption, the government in the future will be able to use this as an excuse for setting up the market in rak being forced by the 2004 withdrawal. " if it doesn't change it can
\end{scriptsize}

\end{tabularx}

\end{table*}

\begin{table*}[!ht]
\caption{\name-on generated text for a SlimPajama-6B-Validation sample with the lowest \name-on perplexity.}
\label{tab:continuation-on-slimpajama-onlybestgeneration}

\begin{tabularx}{1\textwidth}{p{0.43\textwidth}p{0.235\textwidth}p{0.255\textwidth}}
\hline
Context                                                                                                                                                                                                                                                                                                                                                                                                                                                                                                                                                  & Real Continual Chunk {[}1.0{]}                                                                                                                                                                                                                                                        & Multinomial {[}0.594{]}                                                                                                                                                                                                                                                                                  \\ \hline
\begin{scriptsize} (...) it supports tv 3d and 4k, with a resolution of up to 3840x 2160 pixels. audio return channel technology (arc). high quality video and audio hdmi cable with 3 m of length. this cable reaches a data transfer speed of up to 100 mbps and it is hec compatible, so it allows for 2-way communication between devices. it supports tv 3d and 4k, with a resolution of up to 3840 x 2160 pixels. audio return channel technology (arc). high quality video and audio hdmi cable with 4.5 m of length. this cable reaches a data transfer speed of up to 100 mbps and it is hec compatible, so it allows for \end{scriptsize}
& 
\begin{scriptsize}2-way communication between devices. it supports tv 3d and 4k, with a resolution of up to 3840 x 2160 pixels. audio return channel technology (arc). high quality video and audio hdmi cable with 7.5 m of length. this cable reaches a data transfer speed of up to 100 \end{scriptsize}
& 
\begin{scriptsize}the audio transmission speed is 1080p64- hdmi to 720p64 requires all x4 and 1080p64 / 1080p64 / 1080p64 and 1080p64are applied before you check the volume of the rom predict the screen switching machine ispurposed, its operation basically from the udx networks\end{scriptsize}
\\
\\
\end{tabularx}

\end{table*}

\subsubsection{Qualitative Results}
In order to better understand the quantitative results, we present here some generated samples from our worst-performing domain, SlimPajama. We evaluate the semantic similarity of the model-generated chunk to the real continuation chunk. 

For \name-on and \name-off we pick their best-performing validation samples on SlimPajama-Validation, feed 75\% of the context window into the model and generate the next chunk. In order to evaluate the similarity of the generated chunk to the real continuation, we embed it using \textsc{SBert} and report the similarity measure. We employed several generation modes of which we only present the best-performing ones from retrieval on and off in Tables~\ref{tab:continuation-off-slimpajama-onlybestgeneration} and~\ref{tab:continuation-on-slimpajama-onlybestgeneration} respectively, full results for this experiment can be found in Appendix~\ref{app:qualitative_cmp}. 

The dot product results are in brackets for each chunk. We see that the \name-on generated sample is not only more coherent but more similar semantically to the real continuation chunk than the \name-off generated sample is to its own real continuation chunk. 

\subsection{Domain Shift Generalization: Fine-tuning}
\label{sec:finetuning}
In addition to the plug-and-play style domain shift generalization experiments, we conduct experiments in which both \name-on (with regularization of Gaussian with $\lambda_t = 0.2$) and \name-off models undergo a minimal amount of fine-tuning steps with partial parameter updates while the bulk of their parameters remain frozen. The main motivation for this exercise is to assess how far the perplexity performance can go down with a minimalistic fine-tuning effort (in terms of both the number of updated parameters and steps of updates) compared to the plug-and-play domain shift experiments which involved no fine-tuning whatsoever. 

From a selected dataset (e.g. B, C, D), we first randomly sample 15\% of the training set data and reserve it as the input sequences to be used during the fine-tuning phase. The remaining 85\% of the training set data are directed toward building the retrieval database to be used during the fine-tuning phase. Note that the training input sequences and the retrieval database are made mutually exclusive in order to avoid undesired leakage effects during retrieval.

As a starting point, we take the best checkpoints created by training both \name-off and \name-on models on the base WikiText-103 dataset as mentioned in Section~\ref{sec:domainshift_plug_and_play}. For \name-on models we observe updating CCA layer parameters leads to slightly worse performance. Hence we update only the parameters in the feed-forward layers and the read-out layers for both model types during fine-tuning for a few epochs. 

After fine-tuning for a few epochs, the updated model checkpoints are used to run inference on the validation sets from the corresponding datasets. The validation training sequences and the validation retrieval database are prepared exactly the same way as in plug-and-play experiments explained in Section~\ref{sec:domainshift_plug_and_play} such that the performance can be compared under equal settings.

\begin{table*}[!ht]
\centering
\caption{Perplexity results on the validation set for the domain shift experiments of three datasets with fine-tuning the feed-forward and the read-out layers for both \name-on (with the best regularization) and \name-off using different epochs (1--7). Epochs at which the lowest perplexities are achieved are highlighted in each row. The number of training tokens processed per epoch is shown in the \# Train tokens column.}

\label{tab:finetune}
\ra{1.3}
\begin{tabular}{lrlrrrrrrrr}
\hline
\multirow{2}{*}{Dataset}       & \multicolumn{1}{c}{\multirow{2}{*}{\begin{tabular}[c]{@{}c@{}}\# Train\\ tokens\end{tabular}}} & \multirow{2}{*}{Retrieval} & \multicolumn{7}{c}{Fine-tuning epochs}                                                                                                                                            & \multicolumn{1}{l}{}  \\ \cline{4-11} 
                               & \multicolumn{1}{c}{}                                                                          &                            & \multicolumn{1}{c}{0} & \multicolumn{1}{c}{1} & \multicolumn{1}{c}{2} & \multicolumn{1}{c}{3} & \multicolumn{1}{c}{4} & \multicolumn{1}{c}{5} & \multicolumn{1}{c}{6} & \multicolumn{1}{c}{7} \\ \hline
\multirow{2}{*}{BBC-News}      & \multirow{2}{*}{84\,K}                                                                          & Off                        & 143.0                 & 61.1                  & \textbf{57.8}         & 62.2                  & 70.0                  & 84.4                  & 103.7                 & 134.3                 \\
                               &                                                                                               & On ($\lambda_t = 0.2$)                         & 135.6                 & 94.8                  & 79.1                  & 71.4                  & 67.6                  & \textbf{65.3}         & 65.3                  & 66.5                  \\ \hline
\multirow{2}{*}{Reuters}       & \multirow{2}{*}{439\,K}                                                                         & Off                        & 184.6                 & 31.2                  & \textbf{29.6}         & 30.8                  & 33.5                  & 37.7                  & 43.1                  & 53.9                  \\
                               &                                                                                               & On ($\lambda_t = 0.2$)                         & 176.4                 & 39.9                  & 33.9                  & \textbf{31.9}         & 35.6                  & 33.0                  & 35.5                  & 36.8                  \\ \hline
\multirow{2}{*}{CNN-DailyMail} & \multirow{2}{*}{34430\,K}                                                                       & Off                        & 196.0                 & 27.7                  & 26.5                  & \textbf{26.4}         & 26.6                  & 27.3                  & 27.1                  & 27.3                  \\
                               &                                                                                               & On ($\lambda_t = 0.2$)                         & 184.2                 & 29.4                  & 29.1                  & 28.2                  & \textbf{27.9}         & 29.4                  & 30.9                  & 29.7                  \\ \hline
\end{tabular}
\end{table*}

The results of fine-tuning experiments are presented in Table~\ref{tab:finetune}. We evaluate on a subset of datasets used for the plug-and-play domain shift generalization experiments, namely, BBC-News, Reuters, CNN-DailyMail. At epoch=0, i.e., before any fine-tuning, the perplexities reported on the validation show that \name-on consistently dominates \name-off. This is in line with the plug-and-play results in Table~\ref{tab:domain_shift_experiments}; the slight deviations are due to the number of tokens considered in each sequence for the evaluation. 

We observe that for all datasets, the best perplexities are achieved within the first five epochs. This indicates that only a minimal amount of updates are sufficient to fine-tune both \name-on and \name-off models. \name-on variant tends to take one or two additional epochs to reach the best perplexity and remains slightly higher compared to the best perplexity of \name-off. However \name-off shows signs of strong overfitting with training on more and more epochs. In summary, fine-tuning leads to competitive perplexities in both \name-on and \name-off models.

\section{Related Work}
\label{sec:background_and_related_work}

\subsection{RAG and \textsc{Retro}}
\label{sec:retro}
In RAG, there are various approaches to make use of retrieved neighbors. For instance, one approach~\citep{rag} adds the neighbors to enhance the context of an input sequence, whereas \textsc{Retro}~\citep{retro} attends to the neighbors using the CCA mechanism. It is also possible to use both approaches by adding the nearest neighbor to the context window as well as attending to the other neighbors through CCA. \textsc{Retro} outperforms the approaches without CCA slightly on Natural Questions tasks, but~\citep{nvidiaretro} outperforms \textsc{Retro} by a wide margin by combining the two approaches.

Scalability is central for \textsc{Retro} in particular, as it significantly improved results in language modeling once the retrieval database reached one trillion tokens. For the models closer to our model size adding retrieval barely made a difference. The figure on the second page of the \textsc{Retro} paper reveals that although bits-per-byte drops upon adding retrieval, even making the retrieval database almost 10$\times$ larger (up to 10 billion tokens) did not affect the performance.

All of this culminates in a state-of-the-art test perplexity of 2.4 on WikiText-103-Validation. It has to be said that, as the \textsc{Retro}'s authors noted themselves, it is evident that this perplexity is mainly due to validation set leakage. This is practically unavoidable with a retrieval database of this size no matter the validation set, as there is only so much data that can be web-scraped and is public domain.

Furthermore, a retrieval database consisting of trillions of tokens is unrealistic for most setups. Although it is a one-time cost and adding new entries is straightforward, it is difficult to find open-source datasets of this size, let alone have the resources to clean up and process them appropriately.

Recent works~\citep{atlas,raml,mueller2023meta} apply principles of RAG for few-shot learning tasks where retrieval helps with the meta-learning process. However, these models are typically larger compared to \name, operate on different tasks/datasets, and crucially fail to establish a retrieval-off baseline for their respective models. This makes the quantitative comparison of \name against these models somewhat unrealistic.

\subsection{NEFTune} 
\label{sec:neftune}
Harnessing the inherent capabilities of language models for specific tasks can be done in many ways. One is through classical fine-tuning, where we add a task-specific head to a pre-trained model to obtain a task-specific model. We can also use pre-trained models via few-shot prompting for tasks such as generative question answering. Finally, in~\citep{neftune} they settled on instruction tuning, where existing samples were augmented with instructions in natural language. The authors observed that adding uniform noise, reminiscent of noise added in adversarial literature, to the word embeddings of the instructions improved the performance of such models significantly through a regularization effect. Given that one objective of these RAGs is to improve generalization through the non-parametric memory, we investigated adding such a regularizer to our non-parametric memory.
\section{Conclusion and Outlook}
\label{sec:conclusion_and_outlook}
With \name, we have shown that by proper architectural enhancements and training strategies, there is a role for retrieval in the medium-size parametric models using small-sized retrieval databases that are orders of magnitude smaller (570\,K up to 2.89\,B database tokens) compared to \textsc{Retro}. \name consistently improves language modeling and cross-domain generalization compared to the same architecture without retrieval. By applying regularization to the non-parametric memory, we improved the generalization ability of \name even further. Additionally, we have shown that the non-parametric memory can be made robust against noisy similarity searches, which makes it amenable for deployment on the efficient IMC hardware without performance loss. When needed, \name can similarly benefit from fine-tuning.

Future work would evaluate task-specific performance, especially in the domain of question answering. \textsc{Retro} has been shown to lower hallucinations and be less toxic when compared to its non-retrieval counterpart by retrieving from trillions of tokens, but not yet at a small scale. 
Moreover, there are embedding models that are better suited for semantic similarity search than \textsc{SBert}, as measured by the MTEB benchmark. Changing the embedding model is straightforward, although the retrieval database must be re-computed. 

In \name, we attend to all retrieved neighbors indiscriminately, whereas future work could add a similar mechanism as Self-RAG~\citep{selfrag} to decide when and if to retrieve neighbors, or select an optimal subset of them~\citep{RetWhenNeeded}. It might also explore \textsc{Retro}-fitting better foundational models, or different attention-based architectures altogether. 

Future work would also include more accurate hardware-aware training, obtained by applying training techniques similar to the ones presented in~\citep{hwa_imc}. These techniques so far are applied to parametric weights. It will be interesting to see how these training dynamics play out in the non-parametric memories similar to the ones used in \name.

\section*{Acknowledgment}
This work is supported by the Swiss National Science foundation (SNF), grant 200800.


\bibliography{mybibfile}

\begin{thebibliography}{55}
\providecommand{\natexlab}[1]{#1}
\providecommand{\url}[1]{\texttt{#1}}
\expandafter\ifx\csname urlstyle\endcsname\relax
  \providecommand{\doi}[1]{doi: #1}\else
  \providecommand{\doi}{doi: \begingroup \urlstyle{rm}\Url}\fi

\bibitem[Asai et~al.(2024)Asai, Wu, Wang, Sil, and Hajishirzi]{selfrag}
A.~Asai, Z.~Wu, Y.~Wang, A.~Sil, and H.~Hajishirzi.
\newblock Self-{RAG}: Learning to retrieve, generate, and critique through self-reflection.
\newblock In \emph{The Twelfth International Conference on Learning Representations}, 2024.

\bibitem[Borgeaud et~al.(2022)Borgeaud, Mensch, et~al.]{retro}
S.~Borgeaud, A.~Mensch, et~al.
\newblock Improving language models by retrieving from trillions of tokens.
\newblock In \emph{International conference on machine learning}, pages 2206--2240. PMLR, 2022.

\bibitem[Brown et~al.(2020)Brown, Mann, et~al.]{gpt3}
T.~Brown, B.~Mann, et~al.
\newblock Language models are few-shot learners.
\newblock \emph{Advances in neural information processing systems}, 33:\penalty0 1877--1901, 2020.

\bibitem[Computer(2023)]{redpajama}
T.~Computer.
\newblock Redpajama: an open dataset for training large language models, October 2023.
\newblock URL \url{https://github.com/togethercomputer/RedPajama-Data}.

\bibitem[Devlin et~al.(2018)Devlin, Chang, et~al.]{bert}
J.~Devlin, M.-W. Chang, et~al.
\newblock Bert: Pre-training of deep bidirectional transformers for language understanding.
\newblock \emph{arXiv preprint arXiv:1810.04805}, 2018.

\bibitem[Gao et~al.(2020)Gao, Biderman, , et~al.]{thepile}
L.~Gao, S.~Biderman, , et~al.
\newblock The pile: An 800gb dataset of diverse text for language modeling.
\newblock \emph{arXiv preprint arXiv:2101.00027}, 2020.

\bibitem[Gehman et~al.(2020)Gehman, Gururangan, et~al.]{gehman-etal-2020-realtoxicityprompts}
S.~Gehman, S.~Gururangan, et~al.
\newblock {R}eal{T}oxicity{P}rompts: Evaluating neural toxic degeneration in language models.
\newblock In \emph{Findings of the Association for Computational Linguistics: EMNLP 2020}, pages 3356--3369, Nov. 2020.

\bibitem[Gokaslan et~al.(2019)Gokaslan, Cohen, et~al.]{openwebtext}
A.~Gokaslan, V.~Cohen, et~al.
\newblock Openwebtext corpus.
\newblock \url{http://Skylion007.github.io/OpenWebTextCorpus}, 2019.

\bibitem[Greene and Cunningham(2006)]{bbc-news}
D.~Greene and P.~Cunningham.
\newblock Practical solutions to the problem of diagonal dominance in kernel document clustering.
\newblock In \emph{Proc. 23rd International Conference on Machine learning (ICML'06)}, pages 377--384. ACM Press, 2006.

\bibitem[Guo et~al.(2020)Guo, Sun, et~al.]{scann}
R.~Guo, P.~Sun, et~al.
\newblock Accelerating large-scale inference with anisotropic vector quantization.
\newblock In \emph{International Conference on Machine Learning}, pages 3887--3896. PMLR, 2020.

\bibitem[Hayes(2014)]{reuters}
P.~J. Hayes.
\newblock Intelligent high-volume text processing using shallow, domain-specific techniques.
\newblock In \emph{Text-based intelligent systems}, pages 227--241. Psychology Press, 2014.

\bibitem[Henderson et~al.(2022)Henderson, Krass, et~al.]{pileoflaw}
P.~Henderson, M.~Krass, et~al.
\newblock Pile of law: Learning responsible data filtering from the law and a 256gb open-source legal dataset.
\newblock \emph{Advances in Neural Information Processing Systems}, 35:\penalty0 29217--29234, 2022.

\bibitem[Hendrycks et~al.(2021)Hendrycks, Burns, Chen, and Ball]{atticus}
D.~Hendrycks, C.~Burns, A.~Chen, and S.~Ball.
\newblock {CUAD}: An expert-annotated {NLP} dataset for legal contract review.
\newblock In \emph{Thirty-fifth Conference on Neural Information Processing Systems Datasets and Benchmarks Track (Round 1)}, 2021.
\newblock URL \url{https://openreview.net/forum?id=7l1Ygs3Bamw}.

\bibitem[Hermann et~al.(2015)Hermann, Kocisky, et~al.]{cnn_og}
K.~M. Hermann, T.~Kocisky, et~al.
\newblock Teaching machines to read and comprehend.
\newblock \emph{Advances in neural information processing systems}, 28, 2015.

\bibitem[Hoffmann et~al.(2022)Hoffmann, Borgeaud, et~al.]{chinchilla}
J.~Hoffmann, S.~Borgeaud, et~al.
\newblock Training compute-optimal large language models.
\newblock \emph{arXiv preprint arXiv:2203.15556}, 2022.

\bibitem[Izacard and Grave(2021)]{izacard2021leveraging}
G.~Izacard and E.~Grave.
\newblock Leveraging passage retrieval with generative models for open domain question answering.
\newblock In \emph{EACL 2021-16th Conference of the European Chapter of the Association for Computational Linguistics}, pages 874--880. Association for Computational Linguistics, 2021.

\bibitem[Izacard et~al.(2023)Izacard, Lewis, Lomeli, Hosseini, Petroni, Schick, Dwivedi-Yu, Joulin, Riedel, and Grave]{atlas}
G.~Izacard, P.~Lewis, M.~Lomeli, L.~Hosseini, F.~Petroni, T.~Schick, J.~Dwivedi-Yu, A.~Joulin, S.~Riedel, and E.~Grave.
\newblock Atlas: Few-shot learning with retrieval augmented language models.
\newblock \emph{Journal of Machine Learning Research}, 24\penalty0 (251):\penalty0 1--43, 2023.

\bibitem[Jain et~al.(2024)Jain, yeh Chiang, et~al.]{neftune}
N.~Jain, P.~yeh Chiang, et~al.
\newblock {NEFT}une: Noisy embeddings improve instruction finetuning.
\newblock In \emph{The Twelfth International Conference on Learning Representations}, 2024.

\bibitem[Johnson et~al.(2019)Johnson, Douze, and J{\'e}gou]{faiss}
J.~Johnson, M.~Douze, and H.~J{\'e}gou.
\newblock Billion-scale similarity search with gpus.
\newblock \emph{IEEE Transactions on Big Data}, 7\penalty0 (3):\penalty0 535--547, 2019.

\bibitem[Kaplan et~al.(2020)Kaplan, McCandlish, et~al.]{scalinglaw}
J.~Kaplan, S.~McCandlish, et~al.
\newblock Scaling laws for neural language models.
\newblock \emph{arXiv preprint arXiv:2001.08361}, 2020.

\bibitem[Karunaratne et~al.(2021)Karunaratne, Schmuck, Le~Gallo, Cherubini, Benini, Sebastian, and Rahimi]{Karunaratne2021}
G.~Karunaratne, M.~Schmuck, M.~Le~Gallo, G.~Cherubini, L.~Benini, A.~Sebastian, and A.~Rahimi.
\newblock Robust high-dimensional memory-augmented neural networks.
\newblock \emph{Nature Communications}, 12\penalty0 (1):\penalty0 2468, Apr 2021.

\bibitem[Khaddam-Aljameh et~al.(2021)Khaddam-Aljameh, Stanisavljevic, et~al.]{khaddam2021hermes}
R.~Khaddam-Aljameh, M.~Stanisavljevic, et~al.
\newblock Hermes core--a 14nm cmos and {PCM-based} in-memory compute core using an array of {300ps/LSB} linearized {CCO-based ADCs} and local digital processing.
\newblock In \emph{2021 Symposium on VLSI Circuits}, pages 1--2. IEEE, 2021.

\bibitem[Kingma and Ba(2014)]{adam}
D.~P. Kingma and J.~Ba.
\newblock Adam: A method for stochastic optimization.
\newblock \emph{arXiv preprint arXiv:1412.6980}, 2014.

\bibitem[Kudo and Richardson(2018)]{sentencepiece}
T.~Kudo and J.~Richardson.
\newblock Sentencepiece: A simple and language independent subword tokenizer and detokenizer for neural text processing.
\newblock \emph{EMNLP 2018}, page~66, 2018.

\bibitem[Lee et~al.(2022)Lee, Ippolito, et~al.]{lee2022deduplicating}
K.~Lee, D.~Ippolito, et~al.
\newblock Deduplicating training data makes language models better.
\newblock In \emph{Proceedings of the 60th Annual Meeting of the Association for Computational Linguistics (Volume 1: Long Papers)}, pages 8424--8445, 2022.

\bibitem[Lewis et~al.(2020)Lewis, Perez, et~al.]{rag}
P.~Lewis, E.~Perez, et~al.
\newblock Retrieval-augmented generation for knowledge-intensive nlp tasks.
\newblock \emph{Advances in Neural Information Processing Systems}, 33:\penalty0 9459--9474, 2020.

\bibitem[Li et~al.(2023)Li, Li, et~al.]{raml}
R.~Li, Y.~Li, et~al.
\newblock Retrieval-augmented meta learning for low-resource text classification, 2023.

\bibitem[Liu et~al.(2020)Liu, Jiang, et~al.]{radam}
L.~Liu, H.~Jiang, et~al.
\newblock On the variance of the adaptive learning rate and beyond.
\newblock In \emph{International Conference on Learning Representations}, 2020.

\bibitem[Loshchilov and Hutter(2019)]{adamw}
I.~Loshchilov and F.~Hutter.
\newblock Decoupled weight decay regularization.
\newblock In \emph{International Conference on Learning Representations}, 2019.

\bibitem[Merity et~al.(2016)Merity, Xiong, et~al.]{wikitext}
S.~Merity, C.~Xiong, et~al.
\newblock Pointer sentinel mixture models.
\newblock \emph{arXiv preprint arXiv:1609.07843}, 2016.

\bibitem[Mueller et~al.(2023)Mueller, Narang, et~al.]{mueller2023meta}
A.~Mueller, K.~Narang, et~al.
\newblock Meta-training with demonstration retrieval for efficient few-shot learning.
\newblock In \emph{Findings of the Association for Computational Linguistics: ACL 2023}, pages 6049--6064, 2023.

\bibitem[Muennighoff et~al.(2023)Muennighoff, Tazi, Magne, and Reimers]{muennighoff2023mteb}
N.~Muennighoff, N.~Tazi, L.~Magne, and N.~Reimers.
\newblock Mteb: Massive text embedding benchmark.
\newblock In \emph{Proceedings of the 17th Conference of the European Chapter of the Association for Computational Linguistics}, pages 2014--2037, 2023.

\bibitem[Murphy(1996)]{jaccard}
A.~H. Murphy.
\newblock The finley affair: A signal event in the history of forecast verification.
\newblock \emph{Weather and Forecasting}, 11\penalty0 (1):\penalty0 3 -- 20, 1996.

\bibitem[Nallapati et~al.(2016)Nallapati, Zhou, et~al.]{cnn_dailymail}
R.~Nallapati, B.~Zhou, et~al.
\newblock Abstractive text summarization using sequence-to-sequence rnns and beyond.
\newblock \emph{arXiv preprint arXiv:1602.06023}, 2016.

\bibitem[Nandakumar et~al.(2019)Nandakumar, Boybat, Joshi, Piveteau, Le~Gallo, Rajendran, Sebastian, and Eleftheriou]{pcm_noise_model}
S.~R. Nandakumar, I.~Boybat, V.~Joshi, C.~Piveteau, M.~Le~Gallo, B.~Rajendran, A.~Sebastian, and E.~Eleftheriou.
\newblock Phase-change memory models for deep learning training and inference.
\newblock In \emph{2019 26th IEEE International Conference on Electronics, Circuits and Systems (ICECS)}, pages 727--730, 2019.
\newblock \doi{10.1109/ICECS46596.2019.8964852}.

\bibitem[OpenAI et~al.(2024)OpenAI, Achiam, , et~al.]{gpt4}
OpenAI, J.~Achiam, , et~al.
\newblock Gpt-4 technical report, 2024.

\bibitem[Rae et~al.(2021)Rae, Borgeaud, et~al.]{gopher}
J.~W. Rae, S.~Borgeaud, et~al.
\newblock Scaling language models: Methods, analysis \& insights from training gopher.
\newblock \emph{arXiv preprint arXiv:2112.11446}, 2021.

\bibitem[Raffel et~al.(2020)Raffel, Shazeer, et~al.]{c4}
C.~Raffel, N.~Shazeer, et~al.
\newblock Exploring the limits of transfer learning with a unified text-to-text transformer.
\newblock \emph{Journal of machine learning research}, 21\penalty0 (140):\penalty0 1--67, 2020.

\bibitem[Rasch et~al.(2021)Rasch, Moreda, Gokmen, Le~Gallo, Carta, Goldberg, El~Maghraoui, Sebastian, and Narayanan]{analogai-hwkit}
M.~J. Rasch, D.~Moreda, T.~Gokmen, M.~Le~Gallo, F.~Carta, C.~Goldberg, K.~El~Maghraoui, A.~Sebastian, and V.~Narayanan.
\newblock A flexible and fast {PyTorch} toolkit for simulating training and inference on analog crossbar arrays.
\newblock In \emph{2021 IEEE 3rd international conference on artificial intelligence circuits and systems (AICAS)}, pages 1--4. IEEE, 2021.

\bibitem[Rasch et~al.(2023)Rasch, Mackin, et~al.]{hwa_imc}
M.~J. Rasch, C.~Mackin, et~al.
\newblock Hardware-aware training for large-scale and diverse deep learning inference workloads using in-memory computing-based accelerators.
\newblock \emph{Nature communications}, 14\penalty0 (1):\penalty0 5282, 2023.

\bibitem[Reimers and Gurevych(2019)]{sentencetransformer}
N.~Reimers and I.~Gurevych.
\newblock Sentence-bert: Sentence embeddings using siamese bert-networks.
\newblock \emph{arXiv preprint arXiv:1908.10084}, 2019.

\bibitem[Schick et~al.(2024)Schick, Dwivedi-Yu, , et~al.]{toolformer}
T.~Schick, J.~Dwivedi-Yu, , et~al.
\newblock Toolformer: Language models can teach themselves to use tools.
\newblock \emph{Advances in Neural Information Processing Systems}, 36, 2024.

\bibitem[Sebastian et~al.(2020)Sebastian, Le~Gallo, et~al.]{IMC_NatureNano2020}
A.~Sebastian, M.~Le~Gallo, et~al.
\newblock Memory devices and applications for in-memory computing.
\newblock \emph{Nature Nanotechnology}, 15:\penalty0 529--544, 2020.

\bibitem[Singh et~al.(2021)Singh, Reddy, et~al.]{sachan2021endtoend}
D.~Singh, S.~Reddy, et~al.
\newblock End-to-end training of multi-document reader and retriever for open-domain question answering.
\newblock \emph{Advances in Neural Information Processing Systems}, 34:\penalty0 25968--25981, 2021.

\bibitem[Smith et~al.(2022)Smith, Patwary, et~al.]{megatron_training_data}
S.~Smith, M.~Patwary, et~al.
\newblock Using deepspeed and megatron to train megatron-turing nlg 530b, a large-scale generative language model.
\newblock \emph{arXiv preprint arXiv:2201.11990}, 2022.

\bibitem[Soboleva et~al.(2023)Soboleva, Al-Khateeb, et~al.]{slimpajama}
D.~Soboleva, F.~Al-Khateeb, et~al.
\newblock {SlimPajama: A 627B token cleaned and deduplicated version of RedPajama}, June 2023.

\bibitem[Su et~al.(2024)Su, Ahmed, et~al.]{rope}
J.~Su, M.~Ahmed, et~al.
\newblock Roformer: Enhanced transformer with rotary position embedding.
\newblock \emph{Neurocomputing}, 568:\penalty0 127063, 2024.

\bibitem[Sun et~al.(2024)Sun, Xu, et~al.]{kgvslm}
K.~Sun, Y.~Xu, et~al.
\newblock Head-to-tail: How knowledgeable are large language models ({LLM}s)? {A}.{K}.{A}. will {LLM}s replace knowledge graphs?
\newblock In K.~Duh, H.~Gomez, and S.~Bethard, editors, \emph{Proceedings of the 2024 Conference of the North American Chapter of the Association for Computational Linguistics: Human Language Technologies (Volume 1: Long Papers)}, pages 311--325, Mexico City, Mexico, June 2024. Association for Computational Linguistics.
\newblock \doi{10.18653/v1/2024.naacl-long.18}.

\bibitem[US National Archives()]{foundersonline}
US National Archives.
\newblock National historical publications and records commission. founders online.
\newblock \url{https://founders.archives.gov/}, 2024.

\bibitem[Varuna~Jayasiri(2020)]{labml}
N.~W. Varuna~Jayasiri.
\newblock labml.ai annotated paper implementations, 2020.
\newblock URL \url{https://nn.labml.ai/}.

\bibitem[Vaswani et~al.(2017)Vaswani, Shazeer, et~al.]{transformer}
A.~Vaswani, N.~Shazeer, et~al.
\newblock Attention is all you need.
\newblock \emph{Advances in neural information processing systems}, 30, 2017.

\bibitem[Wang et~al.(2023)Wang, Ping, et~al.]{nvidiaretro}
B.~Wang, W.~Ping, et~al.
\newblock Shall we pretrain autoregressive language models with retrieval? a comprehensive study.
\newblock In H.~Bouamor, J.~Pino, and K.~Bali, editors, \emph{Proceedings of the 2023 Conference on Empirical Methods in Natural Language Processing}, pages 7763--7786, Singapore, Dec. 2023. Association for Computational Linguistics.
\newblock \doi{10.18653/v1/2023.emnlp-main.482}.

\bibitem[Wang et~al.(2024)Wang, Huang, et~al.]{RetWhenNeeded}
D.~Wang, Q.~Huang, et~al.
\newblock {Retrieve What You Need: A Mutual Learning Framework for Open-domain Question Answering}.
\newblock \emph{Transactions of the Association for Computational Linguistics}, 12:\penalty0 247--263, 2024.

\bibitem[Zhang et~al.(2017)Zhang, Bengio, et~al.]{zhang2017understanding}
C.~Zhang, S.~Bengio, et~al.
\newblock Understanding deep learning requires rethinking generalization.
\newblock In \emph{International Conference on Learning Representations}, 2017.

\bibitem[Zhu et~al.(2019)Zhu, Cheng, et~al.]{zhu2020freelb}
C.~Zhu, Y.~Cheng, et~al.
\newblock Freelb: Enhanced adversarial training for natural language understanding.
\newblock \emph{arXiv preprint arXiv:1909.11764}, 2019.

\end{thebibliography}

\clearpage
\appendix
\onecolumn
\section*{Appendices}
\section{Datasets}
\label{appendix:datasets}
We describe the dataset used in the language modeling experiments and the datasets used for the domain shift experiments. The goal was to use datasets with strong baselines to compare our model to, and which would need minimal pre-processing for our purposes.

\subsection{WikiText}
\label{sec:wikitext}
The WikiText dataset \citep{wikitext} was created to address the issue of there not being a text dataset with long-form content and original punctuation, capitalization, and numbers. As opposed to web-scraped content, the text is more structured and has fewer typos or colloquialisms. This is because the authors employed a measure of quality control by only considering Wikipedia articles, as well as restricting themselves to verified \emph{good} or \textit{featured} articles.

A \emph{good} article has been nominated by a reviewer as well as confirmed by an impartial editor to be "well-written, contain factually accurate and verifiable information, are broad in coverage, neutral in point of view, stable, and illustrated, where possible, by relevant images with suitable copyright licenses."\footnote{https://en.wikipedia.org/wiki/Wikipedia:Good\_articles.} Only about 0.5\% of Wikipedia articles make the cut.

On the other hand, a \textit{featured} article refers to one of the best articles on Wikipedia as decided upon by the editors. The criteria for a \textit{featured} article are stricter than for a \textbf{good} article, and more comprehensive\footnote{https://en.wikipedia.org/wiki/ Wikipedia:Featured\_article\_criteria.}. Such articles make up around 0.09\% of Wikipedia. More details on the processing steps can be found in Section 4.3 of their paper.

We use WikiText-103 which consists of around 103 million training and 254 thousand validation tokens. We use the same database for training and validation and generate it using WikiText-103-Train. Validation and training sequences of the WikiText dataset are disjoint by design. To avoid leakage during training we ensure that the neighbors of the training sequences are never direct continuations.

Finally, we report the 1-gram Jaccard-Similarity \citep{jaccard} of the training sequences and their two nearest neighbors in Figure \ref{fig:js} and the ten nearest neighbors in Figure \ref{fig:js_10} as a way to determine leakage. As is evident in the histograms, for most neighbors we do not get a Jaccard-Similarity of more than 0.2, so limited leakage is assured.

\begin{figure*}[!ht]
\centering
    \begin{subfigure}[b]{0.32\linewidth}
    \includegraphics[width=\linewidth]{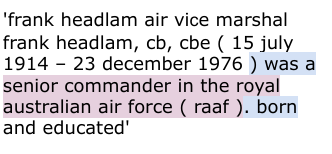}
    \label{fig:ex_chunk}
    \end{subfigure}
    \hfill
    \begin{subfigure}[b]{0.32\linewidth}
    \includegraphics[width=\linewidth]{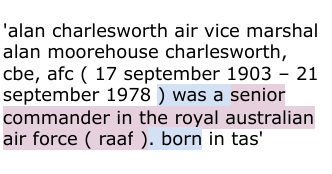}
    \label{fig:ex_n1}
    \end{subfigure}
    \hfill
    \begin{subfigure}[b]{0.32\linewidth}
    \includegraphics[width=\linewidth]{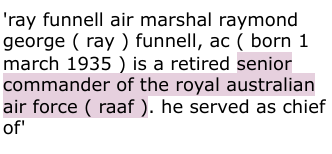}
    \label{fig:ex_n2}
    \end{subfigure}
\caption{Retrieval in action. (a) Extract of the chunk used to query the index (b) Closest neighbor in the retrieval database (c) Second closest neighbor in the retrieval database.}
\label{fig:8t}
\end{figure*}

\begin{figure}[ht]
\centering
    \begin{subfigure}[b]{0.49\linewidth}
    \includegraphics[width=\linewidth]{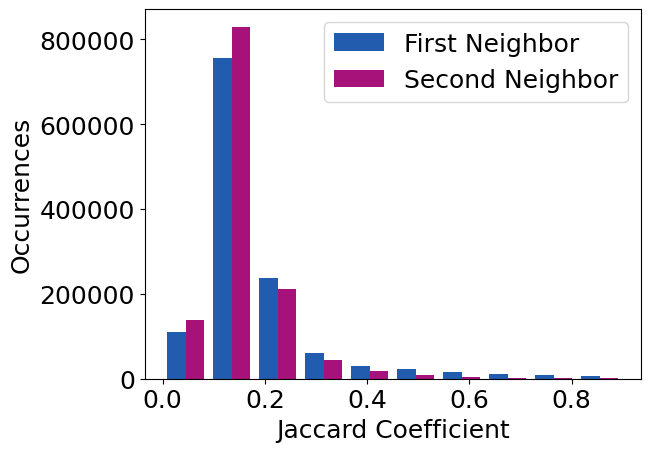}
    \label{fig:js_train}
    \end{subfigure}
    \hfill
    \begin{subfigure}[b]{0.465\linewidth}
    \includegraphics[width=\linewidth]{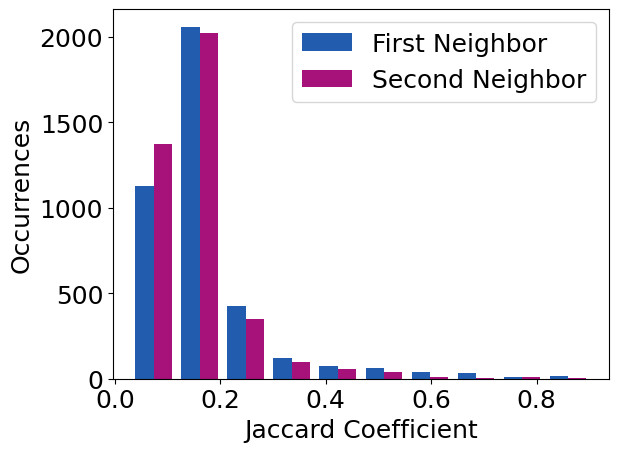}
    \label{fig:js_val}
    \end{subfigure}
\caption{One-gram Jaccard similarity of sequences and their two nearest neighbors (a) For WikiText-103-Train sequences (b) For WikiText-103-Validation sequences.}
\vspace{0.5cm}
\label{fig:js}
\end{figure}

\begin{figure}[H]
    \begin{subfigure}[b]{0.49\textwidth}
    \includegraphics[width=\textwidth]{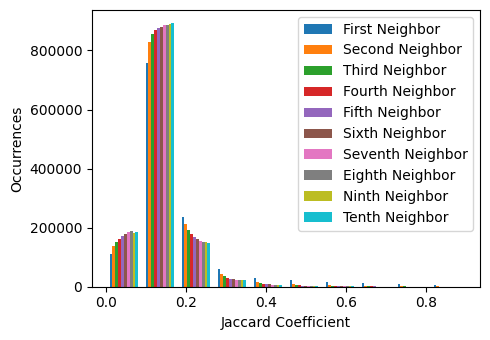}
    \label{fig:js_train-10}
    \end{subfigure}
    \begin{subfigure}[b]{0.49\textwidth}
    \includegraphics[width=\textwidth]{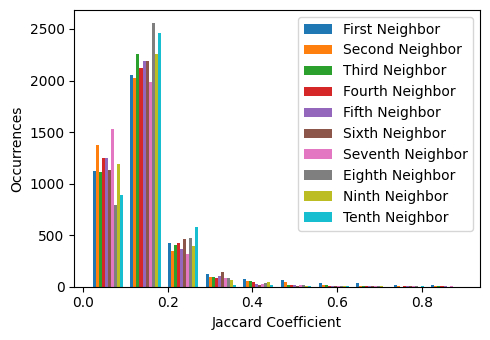}
    \label{fig:js_val-10}
    \end{subfigure}
\caption{One-gram Jaccard similarity of sequences and their ten nearest neighbors (a) For WikiText-103-Train sequences (b) For WikiText-103-Validation sequences.}
\vspace{0.5cm}
\label{fig:js_10}
\end{figure}

In the \textsc{Retro} paper they checked if the training sequences have eight or more contiguous tokens in common with one of their neighbors to determine the degree of leakage. In our case, for the training and validation set, 4.76\% and 5.1\% of the ten nearest neighbors respectively have at least eight contiguous tokens in common with the sequence used to query the index. Evaluating such samples qualitatively, in Figure \ref{fig:8t} we can show that this is hardly leakage in a sense that our model could exploit, as the neighbors come from different articles.

\subsection{Domain shift datasets}
For our domain shift experiments, we choose datasets of varying textual structures and sources. If a train/test/validation split is given, we take the training data to create the retrieval database and validation data to evaluate on. If only a train/test split is given, we use the test data to evaluate on.

\label{app:full_analysis_on_overlap}
Since for the inference experiments we created the retrieval database fully out of the training data and only predicted on the validation data, no leakage is possible.

\subsection{SlimPajama}
\label{app:slimpajama}

\textsc{Retro} was trained on MassiveText \citep{gopher} which is unfortunately proprietary. Furthermore, The Pile \citep{thepile} which is an open-source alternative to MassiveText and on which \textsc{Retro} was also trained for comparison purposes, was taken down in July of 2023 due to a DMCA (Digital Millenium Copyright Act) notice\footnote{https://academictorrents.com/details/ 0d366035664fdf51cfbe9f733953ba325776e667.}. The \textsc{Nvidia} \textsc{Retro} implementation training data was mostly based on The Pile \citep{megatron_training_data} as well. This means that we could not use the same dataset as \textsc{Retro}, which would make it harder for us to compare our performance to theirs. As an alternative, looking for datasets similar in data sources and size to The Pile, we decided on SlimPajama \citep{slimpajama} which is a cleaned and de-duplicated version of RedPajama \citep{redpajama}. De-duplication is especially important for textual training data, see \citep{lee2022deduplicating}. In Table \ref{tab:slimpajamavsthepile} we compare The Pile to SlimPajama, where Pile-CC is comparable to CommonCrawl and C4 \citep{c4}. We see that although the data sources themselves are similar, the proportions are only somewhat comparable.

As most text in the SlimPajama dataset comes from CommonCrawl, it is largely unstructured and conversational. Although for humans it is easy to read and understand, this type of content is significantly dissimilar to Wikipedia articles, in flow, grammar, as well as vocabulary.

We use a sub-sampled version of Slimpajama-627B, namely Slimpajama-6B\footnote{https://huggingface.co/datasets/DKYoon/SlimPajama-6B.}. Furthermore, Slimpajama-6B still needed some clean-up, as there were many instances of repeating characters used as filler or for code comments in the CommonCrawl portion of the dataset. This leads to a total of 5.5B tokens usable for training. 
Finally, to reduce memory usage and simplify the code structure, we truncate all samples longer than 1024 tokens. Considering all of this as well as using their train/validation split, we end up with 2’886’850’140 training tokens and 4’948’252 validation tokens.
\begin{table}[ht]
\centering
\caption{Dataset proportions SlimPajama and The Pile.}
\vspace{0.5cm}
\label{tab:slimpajamavsthepile}
\ra{1.3}
\begin{tabular}{lll}
\toprule
Data source   & SlimPajama & The Pile \\
\midrule
CommonCrawl   & 52.20\%     & 0.00\% \\
C4            & 26.70\%     & 0.00\% \\
Pile-CC       & 0.00\%      & 18.11\%  \\
GitHub        & 5.20\%      & 7.59\%   \\
Books         & 4.20\%      & 12.07\%  \\
ArXiv         & 4.60\%      & 8.96\%   \\
Wikipedia     & 3.80\%      & 1.53\%   \\
StackExchange & 3.30\%      & 5.13\%   \\
Miscellaneous & 0.00\%        & 46.61\% \\
\bottomrule
\end{tabular}
\end{table}

\subsection{BBC-News}
\label{app:bbc-news}
The BBC-News dataset \citep{bbc-news} consists of 2'225 documents from the British Broadcast Corporation news website, across topics such as business, entertainment, politics, sports, and tech. News articles are highly structured in terms of flow, grammar, and vocabulary, so this dataset is similar to WikiText in that regard. Using their train/test split and our tokenization scheme, we end up with 589'677 training tokens and 468'831 validation tokens. The validation set size is unusually large with respect to the training set size. Since we used the training set as our retrieval database, this might have hindered performance somewhat.
\subsection{Reuters}
\label{app:reuters}
The Reuters-21578 dataset as created and used by Hayes in \citep{reuters}, consists of 19'043 documents and has 674 categories in total. This dataset is comprised of news stories from the Reuters financial news-wire service in 1987 and just like the BBC-News dataset, is very similar to WikiText in terms of structure and language. However, the differences in vocabulary might be larger here, as the articles are from decades ago. Using Hayes' train/test split and our tokenization scheme, we end up with 3'454'605 training tokens and 174'335 validation tokens. 
\subsection{Pile-of-Law}
\label{app:pile-of-law}
The Pile-of-Law dataset \citep{pileoflaw} was created in order to mitigate the effects of potentially harmful and biased training data such as can be found through web-crawling alone. The 34 data sources range from "legal case opinions and filings", mostly from the U.S.A., to study materials for law exams and were extensively analyzed for toxicity and biases. We use two of their data sources.
\subsubsection{Atticus Contracts}
\label{app:atticus_contracts}
The Atticus contracts subset contains contracts from the Atticus project, specifically the commercial contracts dataset \citep{atticus}. It consists of 510 contracts and was created to train a model to highlight portions of the contract a human should review. Contract language is highly structured and repetitive, but not necessarily similar to Wikipedia or news articles. We now move away from \textit{article} style text into something new entirely. Using their train/test split and our tokenization scheme, we end up with 456'681'888 training tokens and 152'337'169 validation tokens, which for performance reasons we had to sub-sample into 7'605'872 validation tokens.
\subsubsection{Founding Documents}
\label{app:founding_docs}
This subset consists of letters by U.S.A., founders, which were scraped from Founders Online \citep{foundersonline} and consists of 137'883 training and 45'781 validation samples. Although this is in the Pile-of-Law dataset as well, it is not at all similar in structure to the Atticus contracts. Since these are letters, they are structurally more similar to articles than contracts, but as they are letters from the 1800s, the vocabulary and sentence structure are quite dissimilar to contemporary Wikipedia articles. Using their train/test split and our tokenization scheme, we end up with 55'818'926 training tokens and 18'605'985 validation tokens. 
\subsection{OpenWebText}
\label{app:openwebtext}
The original WebText dataset used to train GPT-2 has not been released. However, as the authors have published how to re-create it, it has been repeatedly reconstructed.  OpenWebText is created by taking the text of articles linked on Reddit\footnote{www.reddit.com.} which have at least 3 upvotes. We use \citep{openwebtext}, this version has over 8 million documents and no inherent train/test split. To keep the sizes manageable, we sub-sample 55\% of the dataset to get 2’491’806’520 tokens which we split by taking 1\% of it for validation. This is large enough to give us a good picture but not so large as to overwhelm the retrieval database size as in BBC-News. 
Overall we end up with 2'477'892'482 training tokens and 13'914'038 validation tokens. This is our second-largest retrieval database, right after SlimPajama-6B.
\subsection{CNN-DailyMail}
\label{app:cnn_dailymail}
The CNN-DailyMail dataset originally created in \citep{cnn_og} for question answering, has been processed for summarization in \citep{cnn_dailymail}, which is the version we use. Since these are again news articles, the text formality and structure are comparable to Reuters and BBC-News. Using their train/test split and our tokenization scheme, we end up with 223'216'223 training tokens and 10'116'369 validation tokens.

\hfill \break

\begin{table}[ht]
\caption{Comparison of features among different dataset domains}
\label{tab:domain_comparison}
\centering
\begin{tabular}{|l|c|c|c|c|c|c|c|c|}
\hline

  & \rotatebox[x=0pt,y=-14pt]{90}{WikiText} &
  \rotatebox[x=0pt,y=-14pt]{90}{BBC-News} &
  \rotatebox[x=0pt,y=-14pt]{90}{Reuters} &
  \rotatebox[x=0pt,y=-14pt]{90}{\begin{tabular}[c]{@{}l@{}}CNN-\\ DailyMail\end{tabular}}&
  \rotatebox[x=0pt,y=-14pt]{90}{\begin{tabular}[c]{@{}l@{}}Atticus \\ Contracts\end{tabular}} &
  \rotatebox[x=0pt,y=-14pt]{90}{\begin{tabular}[c]{@{}l@{}}Founding \\ docs\end{tabular}}&
  \rotatebox[x=0pt,y=-14pt]{90}{\begin{tabular}[c]{@{}l@{}}Open-\\ WebText\end{tabular}} &
  \rotatebox[x=0pt,y=-14pt]{90}{\begin{tabular}[c]{@{}l@{}}Slim-\\ Pajama\end{tabular}} \\ \hline

Structured flow    & \checkmark        & \checkmark        & \checkmark      & \checkmark             &                  &               & \checkmark           &            \\ \hline
Diverse Vocabulary &          &          & \checkmark      &               &                  &               &             &            \\ \hline
Higly formal       &          &          &        &               & \checkmark                &               &             &            \\ \hline
Historical         &          &          &        &               &                  & \checkmark             &             &            \\ \hline
Less curated       &          &          &        &               &                  &               & \checkmark           & \checkmark          \\ \hline
Less formal        &          &          &        &               &                  &               &             & \checkmark          \\ \hline
\end{tabular}
\end{table}

\clearpage
\newpage
\section{Background and Related Work}
\label{app:background_and_related_work}
This is an extension of Section \ref{sec:background_and_related_work}. In the main body of the work, only the three most important papers are chosen in order to adhere to the page limit. The papers presented here and their concepts were also crucial for this work and for possible future avenues to explore.
\subsection{RAG}
\label{app:rag}
In natural language processing the branch of retrieval augmented generation has been gaining traction for a few years now. Given the tasks language models are expected to fulfill, from simple question answering, to fact-checking, to multi-turn conversations, combining a model's parametric memory with a non-parametric database is the natural next step. It builds upon an earlier idea of knowledge graphs \citep{kgvslm} and enables not only a better understanding of our models' reasoning but also makes it easier to update knowledge. This is difficult at best and virtually impossible at worst with only parametric knowledge. 

In the RAG paper \citep{rag} they did exactly that, where retrieval from the non-parametric memory is jointly learned with sequence generation. The neighbors are split into 100-word Wikipedia passages, as opposed to chunks. Just like \textsc{Retro}, RAG computes cross-attention over the neighbors of tokens/sequences. Moreover, RAG's non-parametric memory only consists of a single Wikipedia dump (December 2018), as opposed to \textsc{Retro} which has many data sources. 

RAG set a new state-of-the-art on many open-domain QA tasks and even where they did not, they got close to the performance of more complex systems. Although since then there have been many systems that have outperformed RAG on open-domain QA, such as $\textsc{EDMR}^2$ \citep{sachan2021endtoend} and \textsc{FiD} \citep{izacard2021leveraging} which were specifically trained for QA. It must be said that \textsc{Retro} does not outperform these models either. Though interestingly, \textsc{Nvidia}'s \textsc{Retro++}, where they add the top-1 neighbor to the context, does outperform them.
\subsection{SentenceTransformers}
SentenceTransformers \citep{sentencetransformer} is a Python framework for embeddings of images and text, specifically sentences. While \textsc{Bert} is a general encoder model, only trained for language modeling, SentenceTransformers (also called \textsc{SBert}) is designed to generate embeddings for entire sentences. The purpose is to encode sentences and improve semantic similarity search. 

It works by adding a simple pooling layer after the \textsc{Bert} embeddings. In this pooling layer word embeddings can be averaged, we can take the max embedding or we can take the embedding of the CLS token, which is used for classification.

In this paper, the authors use Siamese and triplet networks to train  \textsc{Bert}. These neural network architectures are designed to learn embeddings for pairs or triplets of data points in a way that emphasizes their similarities and/or differences. Siamese networks have two subnetworks with shared weights. Each subnetwork takes as input a data point (in our case a sentence) and produces embeddings for those inputs. The objective is then to minimize the distance between similar pairs of inputs and maximize the distance between dissimilar pairs. Triplet networks work analogously, but with an anchor (A), a positive example (P), and a negative example (N). The objective is then to minimize the distance between the anchor and the positive example and maximize the distance between the anchor and the negative example. In the end, we obtain sentence embeddings that carry semantic meaning and can be compared using cosine-similarity.

The pre-trained model we use is called \textit{multi-qa-mpnet-base-dot-v1} and is the best SentenceTranformers model for semantic similarity search. It was trained on question-answer pairs, which is the main use case for retrieval-augmented language models.

\subsection{Self-RAG}
Retrieval may not always be helpful and can in some cases be detrimental to the model's performance. In \citep{selfrag} they explored this question, where they retrieved passages on-demand, as opposed to indiscriminately retrieving a fixed number of neighbors like most retrieval systems do, such as our own.

This decision is made using \emph{reflection tokens}, which allow the language model to critique its own output and decide if it needs retrieval to improve the factuality and overall quality of the generated text.

They evaluated their system on fact verification, multiple-choice reasoning, and a variety of question-answering datasets and setups. For these answers, they not only evaluated the exact match but considered fluency and precision/recall as well. They were able to show that their system outperformed other retrieval models on a majority of their tasks and datasets, achieving significant improvements in some cases.

This is an entirely new approach compared to how most RAG systems currently work. Through their evaluation on a diverse set of tasks, they have shown the need for and role of such reflection. We have observed similar issues in our experiments, so the next step would be to add such a self-reflection mechanism as well.

\clearpage
\newpage
\subsection{Chinchilla}
\label{sec:chinchilla}
In \citep{chinchilla} they argue that most large language models are under-trained. The authors analyze the relationship between the number of parameters and training tokens given a fixed compute budget. Given two out of the compute budget, the number of model parameters, and the number of training tokens, they explain and validate through their experiments how to compute the third.

Due to our resource restrictions, we implemented the smallest \textsc{Retro} model with 175M parameters. 
 
Considering Figure 2 in the Chinchilla paper, we can fix the number of parameters and estimate the number of FLOPs as $C = N \cdot D $ where $D$ is the number of training tokens and $N$ is the number of trainable parameters according to \citep{scalinglaw}. For \textsc{Retro-li} with WikiText-103 as both retrieval database and training data, $N = 120M$. We trained on 1'078'012 samples in total, for each sample we have 16 chunks, and for each chunk, we get 2 up to 10 neighbors, where each neighbor is 128 tokens long. This is $5.52\times10^{9}$ to $2.32\times10^{10}$ tokens in total, so C is $3.97\times10^{18}$ to $1.7\times10^{18}$. Even for only two neighbors, this puts us in the optimal model size range.

The non-parametric retrieval database helps with the Chinchilla law, as the retrieved neighbors increase the number of training tokens, alleviating the problem of under-trained large language models.

\section{Regularization}
\label{app:regularization}
We present some details on the NEFTune noise regularization and as it pertains to the approximation to the hardware platform. The regularizer is added to the neighbor embeddings for the CCA blocks during training. For validation and inference, no noise is added to the neighbors, unless specified.

\subsection{Ablation Results}

For this set of experiments, we chose three neighbors and the \textsc{SBert} word embedding model, which gave us the best results so far. Looking at the results for WikiText-103-Validation, we cannot see an improvement of perplexity in any combination of noise and noise placement when averaged over the random seeds, see Tables \ref{tab:neftune_uniform_seq} to  \ref{tab:neftune_uniform_both}. 
\begin{table}[ht]
\caption{Signal to noise ratio averaged over the WikiText-103-Validation word embeddings.}
\label{tab:snr}
\centering
\small
\ra{1.3}
\begin{tabular}{lr}
\toprule
Noise type  & SNR \\
\midrule
Gauss, $\lambda_i=0.4$       & 10.84 \\
Gauss, $\lambda_i=0.2$     & 45.16 \\
Uniform, $\alpha=5$ & 158.49\\
Uniform, $\alpha=10$ & 39.62 \\
Uniform, $\alpha=15$  & 17.61  \\
\bottomrule
\end{tabular}
\end{table}
It is not entirely surprising that the generalization capability of the train to test set did not improve, as adding noise at train time can lead to the model effectively memorizing the noise \citep{zhang2017understanding}. 

The best combination for these experiments is setting $\alpha=10$ and adding noise to the neighbors only, see Table \ref{tab:neftune_noise}. Adding noise to both, sequences and neighbors, see Table \ref{tab:neftune_uniform_both} gives worse results than adding noise only to the sequences. This suggests that the neighbors stabilize the negative effects of the noisy sequence. Furthermore, considering our best result comes from adding noise to the neighbors only, we can see that this NEFTune noise has a regularizing effect on the neighbors. This is likely due to the fact that our retrieval database is too small for our training set. In the original \textsc{Retro} the retrieval database consisted of trillions of tokens and only saw improvement once the size of the retrieval database was in the order of billions, see the Figure on page 1 of the \textsc{Retro} paper as well as the table on page 34.

As we show in Table \ref{tab:domain_shift_experiments_reg}, the checkpoints we trained with uniform noise and $\alpha=10$ do not handle approximation to the hardware platform at inference time any better than the checkpoints trained without noise. Thus, uniform regularization does not play a role in this case. 

\begin{table}[ht]
    \caption{Perplexity results of NEFTune noise with alpha 5, 10, 15 added to sequence, neighbors, and both in \textsc{Retro-li}-on, three neighbors using \textsc{SBert} embeddings, averaged over three random seeds.}
    \vspace{0.5cm}
    \label{tab:neftune_noise}
    \centering
    \ra{1.3}
    \begin{tabular}{llll}
    \toprule
         Alpha  & {Sequence}       & {Neighbors}  & {Both} \\
        \midrule
        {0}   & 22.61          & 22.61  & 22.61 \\
        {5}    & 34.77          & 55.38 & 40.89 \\
        {10}   & 35.94          & 22.77 & 35.96 \\
        {15}  & 34.30          & 50.52  & 36.32 \\
        \bottomrule
    \end{tabular}
\end{table}

\begin{table}[ht]
\centering
\ra{1.3}
\caption{Perplexity results of NEFTune noise added to the training sequence in \textsc{Retro-li}-on, three neighbors, \textsc{SBert} embeddings.}
\vspace{0.5cm}
\label{tab:neftune_uniform_seq}
\begin{subtable}{.49\linewidth}\centering
\caption{Sliding window.}
\label{tab:neftune_uniform_seq_75}
\begin{tabular}{lllll}
\toprule
\begin{tabular}[l]{@{}l@{}}Alpha on\\Sequence\end{tabular}  & 42    & 43    & 44    & avg\\
\midrule
5    & 23.08 & 58.32 & \textbf{22.91} & 34.77\\
10   & 59.68 & \textbf{22.95} & 25.18 & 35.94\\
15   & \textbf{23.02} & 54.68 & 25.20 & \textbf{34.30}\\
\bottomrule
\end{tabular}
\end{subtable}
\begin{subtable}{.49\linewidth}\centering
\caption{Full results.}
\label{tab:neftune_uniform_seq_100}
\begin{tabular}{ccccc}
    \toprule
        \begin{tabular}[l]{@{}l@{}}Alpha on\\Sequence\end{tabular}  & 42    & 43    & 44    & avg\\
        \midrule
        5    & 23.80 & 55.80 & \textbf{23.43} & 34.34\\
        10   & 57.79 & \textbf{23.70} & 25.77 & 35.75\\
        15   & \textbf{23.61} & 52.56 & 24.57 & \textbf{33.87}\\
        \bottomrule
    \end{tabular}
\end{subtable}
\vspace{1cm}
\caption{NEFTune noise added to the neighbors in \textsc{Retro-li}-on, three neighbors, \textsc{SBert} embeddings.}
\vspace{0.5cm}
\label{tab:neftune_uniform_neighbor}
\begin{subtable}{.49\linewidth}\centering
\caption{Sliding window.}
\label{tab:neftune_uniform_neighbor_75}
\begin{tabular}{lllll}
\toprule
\begin{tabular}[l]{@{}l@{}}Alpha on\\Neighbors\end{tabular}& 42    & 43    & 44    & avg\\
\midrule
5    & \textbf{22.23} & 85.07 & 58.83 & 55.38\\
10   & 25.62 & \textbf{21.18} & \textbf{21.52} & \textbf{22.77}\\
15   & 23.96 & 60.31 & 67.30 & 50.52\\
\bottomrule
\end{tabular}
\end{subtable}
\begin{subtable}{.49\linewidth}\centering
\caption{Full results.}
\label{tab:neftune_uniform_neighbor_100}
\begin{tabular}{ccccc}
\toprule
\begin{tabular}[l]{@{}l@{}}Alpha on\\Neighbors\end{tabular}& 42    & 43    & 44    & avg\\
\midrule
5    & \textbf{22.73} & 79.83 & 61.45 & 54.67\\
10   & 26.39 & \textbf{21.63} & \textbf{22.11} & \textbf{23.38}\\
15   & 24.57 & 59.17 & 63.12 & 48.95\\
\bottomrule
\end{tabular}
\end{subtable}
\vspace{1cm}
\caption{NEFTune noise added to both training sequence and neighbors in \textsc{Retro-li}-on, three neighbors, \textsc{SBert} embeddings.}
\vspace{0.5cm}
\label{tab:neftune_uniform_both}
\begin{subtable}{.49\linewidth}\centering
\caption{Sliding window.}
\label{tab:neftune_uniform_both_75}
\begin{tabular}{lllll}
    \toprule
        \begin{tabular}[l]{@{}l@{}}Alpha on\\Both\end{tabular} & 42             & 43             & 44    & avg\\
        \midrule
        5    & \textbf{21.70} & \textbf{22.80} & 78.16 & 40.89\\
        10   & 23.69          & 60.72          & 23.47 & \textbf{35.96}\\
        15   & 25.01          & 61.31          & \textbf{22.65} & 36.32\\
        \bottomrule
    \end{tabular}
\end{subtable}
\begin{subtable}{.49\linewidth}\centering
\caption{Full results.}
\label{tab:neftune_uniform_both_100}
\begin{tabular}{ccccc}
    \toprule
        \begin{tabular}[l]{@{}l@{}}Alpha on\\Both\end{tabular} & 42             & 43             & 44    & avg\\
        \midrule
        5    & \textbf{22.28} & \textbf{23.34} & 72.95 & 39.52\\
        10   & 24.25          & 59.01          & 24.04 & 35.77\\
        15   & 25.62          & 58.26          & \textbf{23.27} & \textbf{35.72}\\
        \bottomrule
    \end{tabular}
\end{subtable}
\end{table}

\begin{figure}
    \centering
    \includegraphics[width=0.7\linewidth]{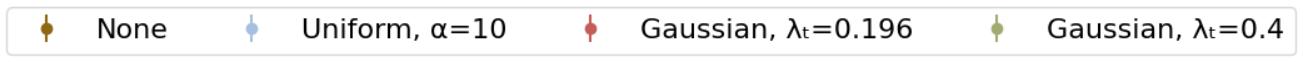}
\begin{subfigure}[b]{0.334\linewidth}
    \includegraphics[width=\linewidth]{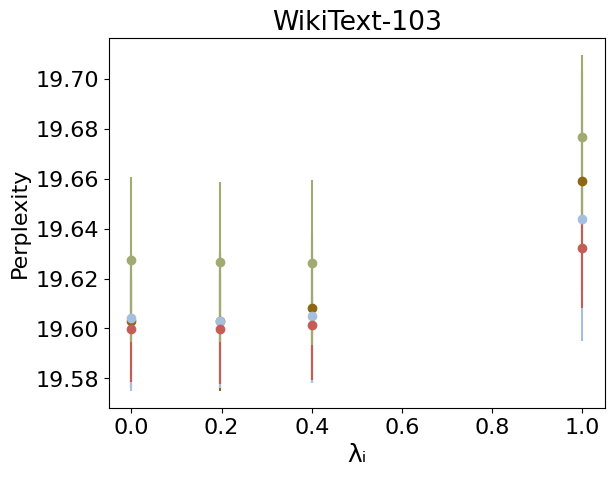}
\end{subfigure}
\begin{subfigure}[b]{0.322\linewidth}
\includegraphics[width=\linewidth]{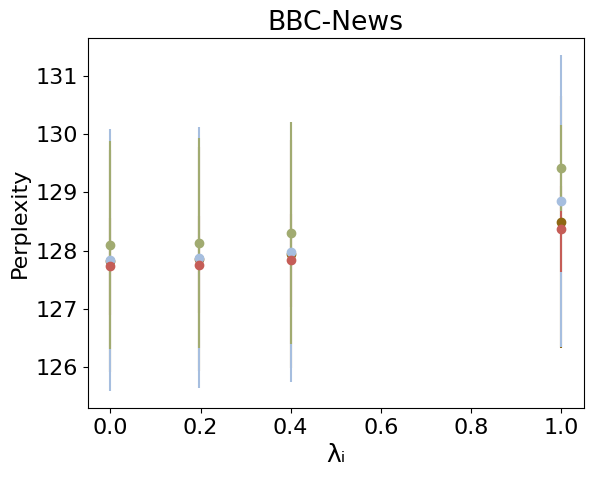}
\end{subfigure}
\begin{subfigure}[b]{0.322\linewidth}
\includegraphics[width=\linewidth]{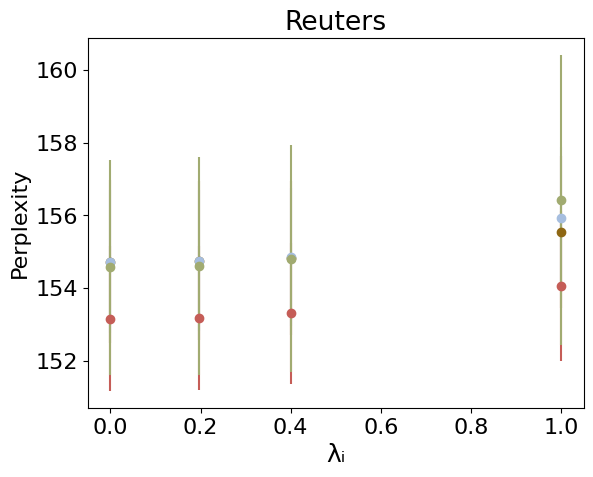}
\end{subfigure}
\begin{subfigure}[b]{0.322\linewidth}
\includegraphics[width=\linewidth]{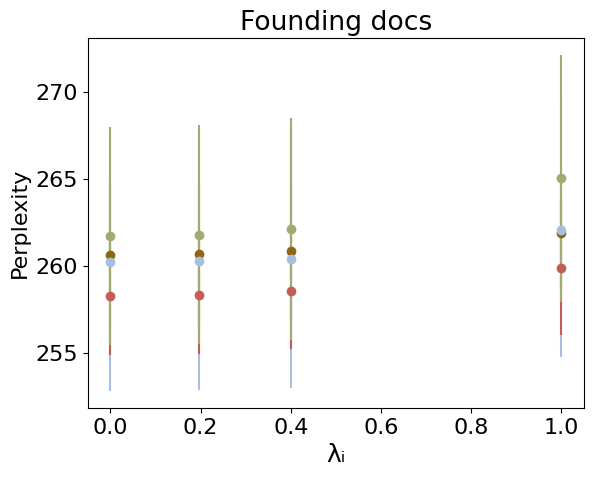}
\end{subfigure}
\begin{subfigure}[b]{0.322\linewidth}
\includegraphics[width=\linewidth]{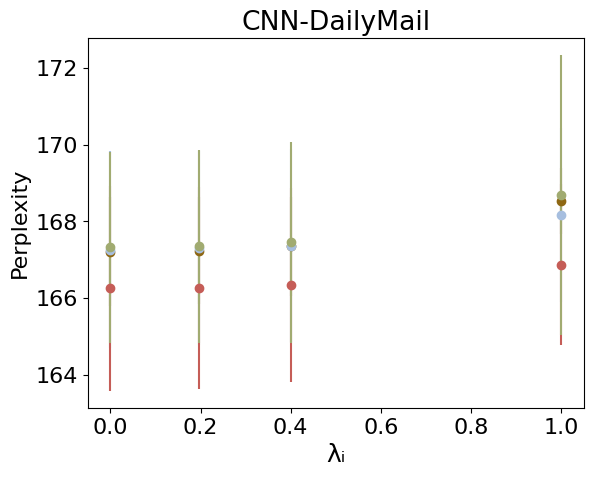}
\end{subfigure}
\begin{subfigure}[b]{0.322\linewidth}
    \includegraphics[width=\linewidth]{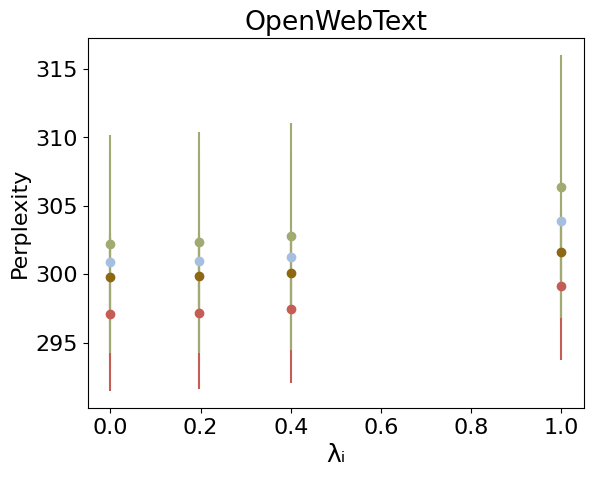}
\end{subfigure}
\begin{subfigure}[b]{0.322\linewidth}
    \includegraphics[width=\linewidth]{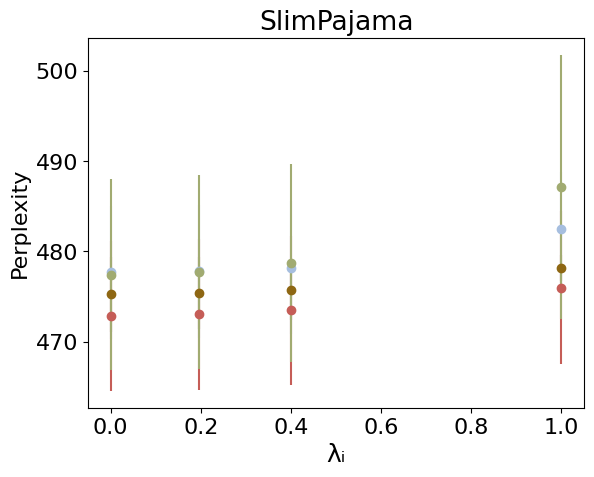}
\end{subfigure}
\caption{Perplexity evolution with increased inference noise for all datasets.}
\vspace{0.5cm}
\label{fig:ppl_per_inferencenoise_all_datasets}
\end{figure}

\begin{table*}[ht]
\caption{Perplexity results for the language modeling and the domain shift experiments, averaged over six random seeds. Models with various settings are evaluated: \name-off and \name-on (without neighbors, with ideal neighbors, and with noisy neighbors impacted by $\lambda_i \in \{0.2,0.4,1\}$).}
\vspace{0.5cm}
\label{tab:domain_shift_experiments_reg}
\centering
\ra{1.3}
\small
\begin{tabular}{@{}lllcccccccc@{}}
\toprule
 & 
   &
   &
   
  \rotatebox[x=0pt,y=-14pt]{90}{WikiText} &
  \rotatebox[x=0pt,y=-14pt]{90}{BBC-News} &
  \rotatebox[x=0pt,y=-14pt]{90}{Reuters} &
  \rotatebox[x=0pt,y=-14pt]{90}{\begin{tabular}[c]{@{}l@{}}Founding \\ docs\end{tabular}}&
  \rotatebox[x=0pt,y=-14pt]{90}{\begin{tabular}[c]{@{}l@{}}CNN-\\ DailyMail\end{tabular}}&
  \rotatebox[x=0pt,y=-14pt]{90}{\begin{tabular}[c]{@{}l@{}}Atticus \\ contracts\end{tabular}} &
  \rotatebox[x=0pt,y=-14pt]{90}{\begin{tabular}[c]{@{}l@{}}Open-\\ WebText\end{tabular}} &
  \rotatebox[x=0pt,y=-14pt]{90}{\begin{tabular}[c]{@{}l@{}}Slim-\\ Pajama\end{tabular}} \\ \cmidrule(l){4-11} Retrieval 
 & Settings &
  Regularizer & & & & & & & & \\ \midrule
off & Normal inf. & None & 20.62 & 130.31 & 168.91 & 266.85 & 171.56 & 329.98 & 312.15 & 489.46 \\ \midrule
\multirow{15}{*}{On} &
  \multirow{4}{*}{\begin{tabular}[c]{@{}l@{}}No neighbors\end{tabular}} &
  None &  19.74 &  129.63 &  157.30 &  265.53 &  170.43 & 322.67 &  306.10 &  485.84 \\
 &   
 &  
 Uniform, $\alpha=10$ &  \textbf{19.71} &  130.51 &  157.79 &  265.96 &  170.51 &  324.63 &  308.61 &  491.11 \\
 &   
 &  
 Gaussian,$\lambda_t=0.2$ &  19.72 & \textbf{129.17}  & \textbf{155.05} & \textbf{260.16} & \textbf{168.78} & \textbf{318.37} & \textbf{301.94} &  \textbf{481.10}\\ 
 &   
 &  
 Gaussian,$\lambda_t=0.4$ & 19.71  & 130.38  & 157.47 & 268.32 & 170.78 & 327.63 & 309.84 & 504.20 \\ \cmidrule(l){2-11} 
 &
  \multirow{4}{*}{Ideal retrieval} &
  None &  19.60 &  127.82 &  154.73 &  260.63 &  167.21 &  316.48 &  299.81 & 475.23 \\
 &
 &
  Uniform, $\alpha=10$ &  19.60 &  127.84 &  154.73 &  260.23 &  167.26 &  318.01 &  300.87 &  477.70 \\
 &
 &
  Gaussian,$\lambda_t=0.2$ & \textbf{19.60} & \textbf{127.73} & \textbf{153.16} & \textbf{258.27} & \textbf{166.25} & \textbf{314.28} & \textbf{297.08} & \textbf{472.87} \\
 &
 &
 Gaussian,$\lambda_t=0.4$ & 19.63 & 128.10 & 154.57 & 261.70 & 167.32 & 317.94 & 302.20 & 477.42\\ \cmidrule(l){2-11} 
 &
  \multirow{4}{*}{\begin{tabular}[c]{@{}l@{}}Noisy retrieval, \\ $\lambda_i=0.2$ \end{tabular}} &  
  None &  19.60 &  127.86 &  154.75 &  260.68 &  167.24 &  316.54 &  299.88 &  475.36 \\
 &
 &
  Uniform, $\alpha=10$ &  19.60 &  127.88 &  154.76 &  260.26 &  167.30 &  318.17 &  300.99 &  477.79 \\
 &
 &
  Gaussian,$\lambda=0.2$ & \textbf{19.60} & \textbf{127.75} & \textbf{153.18} & \textbf{258.34} & \textbf{166.27} & \textbf{314.35} & \textbf{297.18} & \textbf{473.02} \\
 &
 &
  Gaussian,$\lambda=0.4$ & 19.63 & 128.13 & 154.62 & 261.80 & 167.35 & 318.06 & 302.33 & 477.71\\ \cmidrule(l){2-11} 
 &
  \multirow{4}{*}{\begin{tabular}[c]{@{}l@{}}Noisy retrieval, \\ $\lambda_i=0.4$\end{tabular}} &
  None &  19.61 &  127.94 &  154.84 &  260.86 &  167.35 &  316.83 &  300.10 &  475.76 \\
 &
 &
  Uniform, $\alpha=10$&  19.61 &  127.97 &  154.87 &  260.41 &  167.37 &  318.39 &  301.24 &  478.18 \\
 &
 &
  Gaussian, $\lambda_t=0.2$& \textbf{19.60} & \textbf{127.83} & \textbf{153.31} & \textbf{258.58} & \textbf{166.35} & \textbf{314.60} & \textbf{297.47} & \textbf{473.45} \\
 &
 &
  Gaussian, $\lambda_t=0.4$& 19.63 & 128.30 & 154.81 & 262.11 & 167.46 & 318.44 & 302.76 & 478.73 \\ \cmidrule(l){2-11} 
 &
  \multirow{4}{*}{\begin{tabular}[c]{@{}l@{}}Noisy retrieval, \\ $\lambda_i=1.0$\end{tabular}} &
  None &  19.66 &  128.49 &  155.54 &  261.92 &  168.53 &  318.66 &  301.63 &  478.18 \\
 &
 &
  Uniform, $\alpha=10$ &  19.64 &  128.85 &  155.94 &  262.07 &  168.17 &  321.32 &  303.92 &  482.51 \\
 &
 &
  Gaussian, $\lambda_t=0.2$ & \textbf{19.63} & \textbf{128.36} & \textbf{154.06} & \textbf{259.88} & \textbf{166.87} & \textbf{316.00} & \textbf{299.16} & \textbf{475.97}\\
 &
 &
  Gaussian, $\lambda_t=0.4$ & 19.68 & 129.42 & 156.42 & 265.03 & 168.69 & 322.21 & 306.40 & 504.20\\
  \bottomrule
\end{tabular}
\end{table*}
\clearpage
It is clear in Table \ref{tab:domain_shift_experiments_reg} and Figures \ref{fig:ppl_per_inferencenoise_all_datasets} that not only is \textit{Gaussian, $\lambda_t=0.2$} the best checkpoint overall it can handle large amounts of noise better than other regularizers or no regularizer. We see how for $\lambda_i=0.2$ and $\lambda_i=0.4$ the perplexity barely increases for any of the regularizers (None, Uniform, Gaussian). For $\lambda_i=1.0$ which is the largest relative standard deviation in our setup this changes. Here clear patterns emerge. For instance, \textit{Gaussian, $\lambda_t=0.4$} does consistently worst because it has the lowest signal-to-noise ratio of the regularizers shown here. It is most affected by $\lambda_i=1.0$. Moreover, \textit{Gaussian, $\lambda_t=0.2$} does best for every dataset and noise level at inference time. It is least affected by $\lambda_i=1.0$. This is unsurprising, as we also observe that out of all the regularizers \textit{Gaussian, $\lambda_t=0.2$} is least affected by the inference mode \textit{no retrieval}. Most of this model's generalization performance comes from improved language modeling and not from the retrieved neighbors themselves.

The \textit{uniform regularizer} and \textit{no regularizer} get similar results for almost all combinations of datasets and types of inference noise, but that changes for $\lambda_i=1.0$. For all datasets except BBC-News and WikiText-103, \textit{no regularizer} is less affected by $\lambda_i=1.0$ than the \textit{uniform regularizer}. However, it has to be said that, even for these two datasets, the performance decrease is very similar.

\clearpage
\newpage
\section{Perplexity Discussion}
\label{app:perplexity_discussion}
Perplexity is a performance measure in natural language processing (NLP) used to evaluate how well a model predicts a sample. It quantifies how a probability distribution or language model represents the data it is trained on. A lower perplexity score indicates that the model is more accurate and has a better understanding of the data. The model is essentially less surprised by the next word it sees, hence the name. In language modeling, perplexity is the exponential of the cross-entropy loss. In NLP we usually use the exponent base as opposed to base 2 but both are acceptable.

The state-of-the-art language modeling perplexity of 2.4 on WikiText-103-Test is achieved by the \textsc{Retro} model\footnote{https://paperswithcode.com/sota/language-modelling-on-wikitext-103.}, with a large gap to the second best perplexity of 10.6. As already discussed, advancements of perplexity on WikiText-103 are usually due to dataset leakage. Most if not all large language models train on some snapshot of English Wikipedia. With the emergence of foundational models, it is nearly impossible to avoid validation/test set leakage.

Perplexity has other issues as well. As it is a manner of evaluating the probability distribution, it cannot handle words with zero probability, so out of vocabulary words. Language models with a closed vocabulary will have lower perplexity than those with an open vocabulary, which assign probabilities to all words, if need be down to the characters. Finally, it is difficult to compare perplexities between closed vocabulary tokenizers, due to the size of the vocabularies. A larger vocabulary can lead to higher perplexities, as there are more words to consider for the next word.

Despite these issues, perplexity is used as a performance measure due to its simplicity and comparability. In NLP we are aware of these issues and attempt to alleviate them by using closed vocabularies of similar sizes, when our goal is to present and compare perplexities. Furthermore, it is difficult to find a quantitative measure for language models. The BLEU or ROUGE score, \textit{Word error rate}, or \textit{Character error rate} all depend on generating text. Generation is already the next step which is subject to hyperparameter-tuning and post-processing. Not to mention, that the goal might not be to generate the exact continuation, as depending on the use-case, semantically similar output should be accepted as well. To compare the output of the language model directly, we must use a measure that does not sample but evaluates the probability distributions directly.

Recently, bits-per-byte is used in order to measure the performance of a language model. Bits-per-byte refers to data storage efficiency, it describes a compression ratio. The idea behind it is the same as behind perplexity, where $bpb=0$ means that the model knows exactly what symbol will be next whereas $bpb = log_2(vocab\_size)$ means the model needs to be given the next symbol exactly. It is related to perplexity in the sense that cross-entropy loss for a character-level language model averaged over a dataset is bits-per-character. 
It is computed as $L \times log_2(e)$ where $L$ is the loss and is a loss-based performance measure. As it also highly depends on the vocabulary size, it is not a more general measure than perplexity.

\subsection{Full Results}
Here we present the full result tables, for each random seed and all number of neighbors. Variations across random seeds are likely due to the fact that we have to fix the number of training steps. We do not stop once we overfit, as we do not reach that stage with most of our experiments. For full transparency, we present both, the perplexity of the sliding window approach (as \textsc{Retro} computes it) and the full results (simply exp(loss)).
\subsubsection{Embeddings}
We see strong variations across random seeds, suggesting that our system is sensitive to the initial samples and gets stuck in local minima quickly. Moreover, as we do not observe this behavior for \textsc{SBert} embeddings, we can conclude that this is due to less-than-ideal neighbors, which strongly impact the system, especially in the beginning. It has to be said that even \textsc{Retro}-Off has one particularly bad seed, which suggests that at least part of the reason we get stuck in local minima is due to the loss landscape itself.

Finally, note how not all checkpoints improve upon utilizing the sliding window approach. This suggests that this approach is better suited for demonstrating the performance of a language model. If the performance decreases given 75\% of the context, the model is truly unsuited for the task. 
\begin{table}[ht]
\centering
\ra{1.3}
\caption{Perplexity results of experiments for \textsc{Retro-li}-on/-off with \textsc{Bert} embeddings on a variety of number of neighbors and random seeds.}
\vspace{0.5cm}
\begin{subtable}{.50\linewidth}\centering
\caption{Sliding window.}
\label{tab:bert_75}
\begin{tabular}{lllll}
\toprule
\textsc{Bert} & {42} & {43} & {44} & {avg} \\
\midrule
{2}        & 25.51         & 61.17         & 65.11         & 50.59\\
{3}        & 54.61         & 56.90         & 23.29         & 44.93\\
{5}        & 24.24         & 22.67         & 39.85         & \textbf{28.92}\\
{7}        & 26.35         & 60.96         & 60.02         & 49.11\\
{10}       & \textbf{23.48}& 60.21         & 63.62         & 49.10\\
\textsc{Retro-li}-Off& 55.56         & \textbf{21.60}& \textbf{22.02}& 32.06\\
\bottomrule
\end{tabular}
\end{subtable}
\begin{subtable}{.48\linewidth}\centering
\caption{Full results.}
\label{tab:bert_100}
\begin{tabular}{lllll}
\toprule
\textsc{Bert} & {42} & {43} & {44} & {avg} \\
\midrule
{2}        & 26.39         & 61.28         & 66.10      & 51.26\\
{3}        & 52.67         & 53.91         & 23.77      & 43.45\\
{5}        & 24.74         & 23.23         & 39.87      & \textbf{29.28}\\
{7}        & 27.09         & 58.97         & 58.75      & 48.27\\
{10}       & \textbf{24.10}& 57.01         & 61.41      & 47.51\\
\textsc{Retro-li}-Off& 53.17         & \textbf{22.21}& \textbf{22.59}& 32.66\\
\bottomrule
\end{tabular}
\end{subtable}
\end{table}

\begin{table}[ht]
\centering
\caption{Perplexity results of experiments for \textsc{Retro-li}-on/-off with \textsc{SBert} embeddings on a variety of number of neighbors and random seeds.}
\vspace{0.5cm}
\ra{1.3}
\begin{subtable}{.50\linewidth}\centering
\caption{Sliding window.}
\label{tab:sent_75}
\begin{tabular}{lllll}
\toprule
{\textsc{SBert}} & {42} & {43} & {44} & {avg} \\
\midrule
{2}        & 24.26         & 24.07         & 23.82         & 24.05\\
{3}        & \textbf{21.35}& 24.59         & \textbf{21.89}& \textbf{22.61}\\
{5}        & 24.66         & 21.83         & 58.84         & 35.11\\
{7}        & 22.77         & 25.39         & 54.09         & 34.08\\
{10}       & 21.49         & 24.26         & 23.97         & 23.24\\
\textsc{Retro-li}-Off& 55.56         & \textbf{21.60}& 22.02         & 32.06\\
\bottomrule
\end{tabular}
\end{subtable}
\begin{subtable}{.48\linewidth}\centering
\caption{Full results.}
\label{tab:sent_100}
\begin{tabular}{lllll}
\toprule
{\textsc{SBert}} & {42} & {43} & {44} & {avg} \\
\midrule
{2}        & 24.95         & 24.84         & 24.39         & 24.73\\
{3}        & \textbf{22.05}& 25.01         & \textbf{22.52}& \textbf{23.19}\\
{5}        & 25.42         & 22.82         & 65.00         & 37.75\\
{7}        & 23.43         & 26.10         & 56.32         & 35.28\\
{10}       & 22.10         & 25.05         & 24.80         & 23.98\\
\textsc{Retro-li}-Off& 53.17         & \textbf{22.21}& 22.59         & 32.66\\
\bottomrule
\end{tabular}
\end{subtable}
\end{table}

\clearpage
\newpage
\section{Implementing \textsc{Retro}}
\label{app:architecture}
In this section, we provide additional details on the architecture introduced in Section \ref{sec:implementing_retro}. Specifically, we elaborate on other changes we made to the \textsc{Retro} architecture and how we implemented them.

\subsection{\textsc{Retro}-fitted GPT-2}
\label{sec:RetrofittedGPT2}
One contribution of the \textsc{Retro} paper is that this type of architecture is straightforward to apply to other language models. This is called \textsc{Retro}-fitting. This is beneficial and efficient, as it allows us to leverage pre-trained models and augment them with retrieval capabilities. To do so, we must be able to access each attention block separately, as the chunked cross-attention blocks are added to every third layer starting from the sixth (or ninth for larger models). These attention blocks can then be frozen.

This reduces the number of trainable parameters, making it faster to train. In the paper, instead of taking a pre-trained GPT-2 checkpoint, the authors built a GPT-2-like model with the appropriately changed parameters (see Table \ref{tab:gpt2vsretro})
. Then they trained it without retrieval before \textsc{Retro}-fitting it. Due to our limited resources, we use the publicly available GPT-2 checkpoints. Consequently, we had to change the aforementioned hyperparameters. Moreover, we had to train additional feed-forward network parameters at the end of the attention layers. Using a pre-trained model leads to a significant improvement of perplexity on the language modeling task compared to training \textsc{Retro} from scratch and it decreases our model size by 30\%.

\begin{figure}[H]
\centering
    \begin{subfigure}[t]{0.50\linewidth}
    \includegraphics[width=\linewidth]{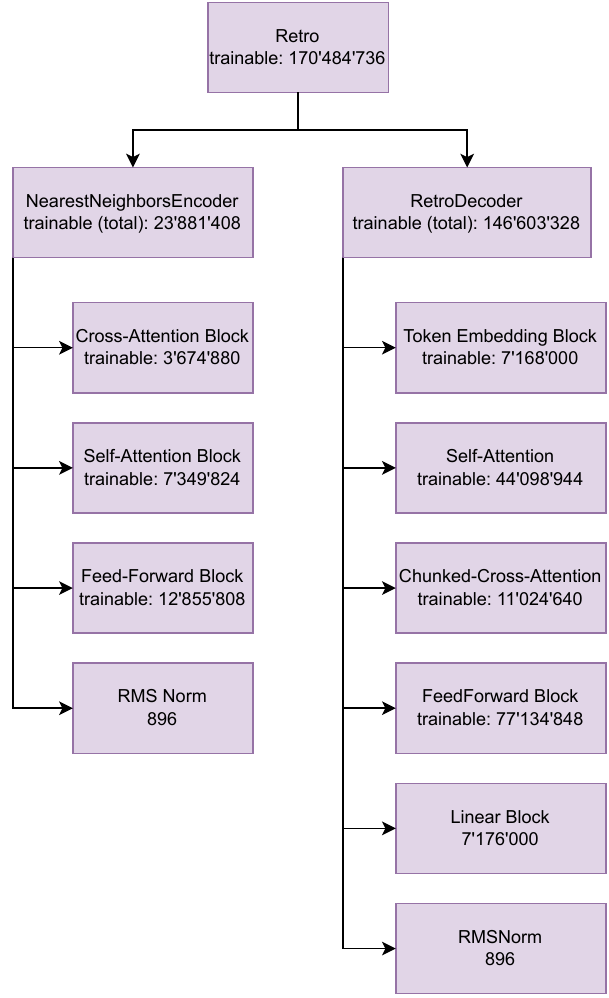}
    \label{fig:retro-og}
    \end{subfigure}
    \hfill
    \begin{subfigure}[t]{0.49\linewidth}
    \includegraphics[width=\linewidth]{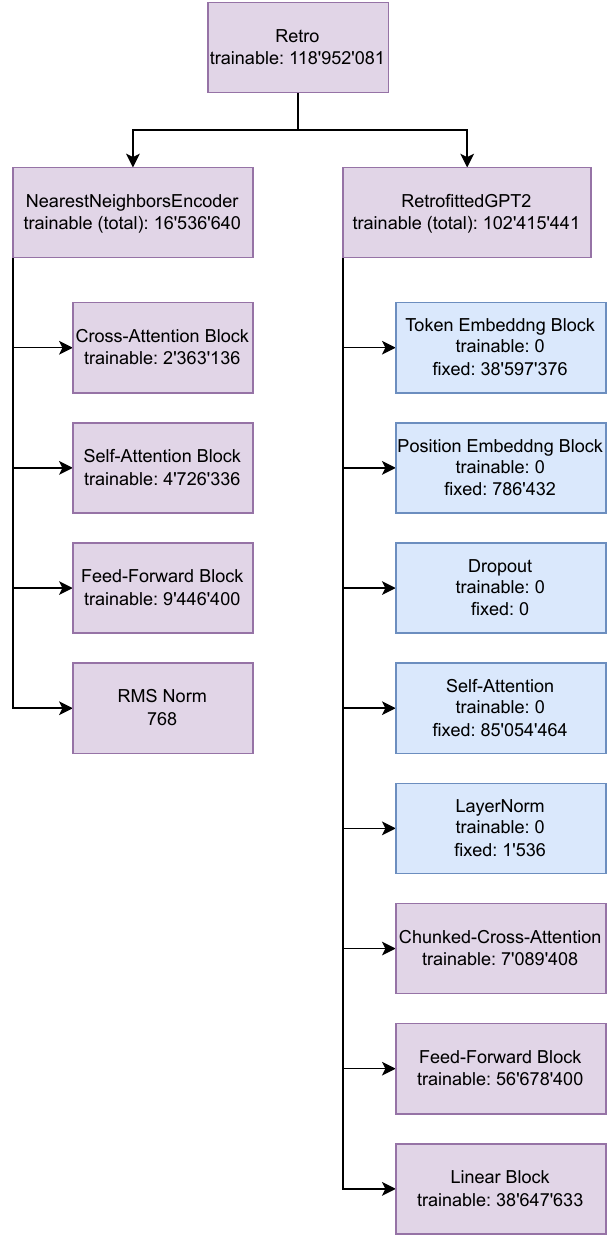}
    \label{fig:retro-gpt}
    \end{subfigure}
\caption{Number of trainable \mbox{(violet)} and frozen (blue) parameters in the (a) \textsc{Retro} architecture and (b) \textsc{Retro}-fitted GPT-2 architecture.}
\label{fig:arch-retro}
\end{figure}

For later experiments, we use our trained \textsc{Retro-li}-off with GPT-2 attention blocks as a checkpoint. This enables us to freeze all decoder layers except the CCA, just as in the original paper. Additionally, it decreases the number of trainable parameters even further from the complete \textsc{Retro-li} by 90.29\%. The architecture is visualized in Diagram \ref{fig:arch-retro}. 

Note that \textsc{Retro} differentiates between the baseline transformer, which is their GPT-2-like model, and \textsc{Retro}-off, which is their \textsc{Retro}-on without CCA. However, they repeatedly demonstrated that their performances are almost identical. In our work, we had no baseline transformer. We compare the \textsc{Retro} architecture with and without CCA.
\begin{table}[ht]
\caption{Changes to the hyperparameters for \textsc{Retro}-fitting.}
\label{tab:gpt2vsretro}
\centering
\ra{1.3}
\begin{tabular}{lll}
\toprule
{Hyperparameter}          & {GPT-2}      & {\textsc{Retro}} \\
\midrule
{d\_model}                & 896        & 768 \\
{d\_ff}                   & 3584       & 3072 \\
{\# attention heads}     & 12         & 16 \\
{vocabulary size}         & 50'257     & 
8'000\tablefootnote{The vocabulary size of the original \textsc{Retro} is 128'000, but our training data is much smaller thus we chose an appropriately smaller vocabulary size.}  \\
\bottomrule
\end{tabular}
\end{table}

\subsection{Changes}
Most changes are made in order to adapt the \textsc{Retro} architecture to the GPT-2 attention blocks.
\subsubsection{Index}
\label{app:index}
\textsc{Faiss} is an open-source library for similarity search created by Facebook Research. Through clustering and quantization, it efficiently searches billions of vectors in milliseconds. Although \textsc{Retro} used ScaNN for their purposes, we chose \textsc{Faiss} due to its ease of use and its parallelization abilities. 

\textsc{Faiss} has several hyperparameters that must be set following the size of the database. According to their own recommendation, the number of centroids\footnote{Also called inverted lists.} should be set as about $\sqrt{n}$ where n is the number of index entries, in our case this refers to the number of chunks. For WikiText-103 this ends up being around 1024, for SlimPajama-6B and OpenWebText, this is around 4096. The number of inverted lists to be searched at query time can be set freely. \textsc{Nvidia}'s Megatron \textsc{Retro} implementation set it as 0.1\% of the inverted lists, whereas we stick to $\sqrt{n_{centroids}}$.

\begin{table}[ht]
\ra{1.4}
\centering
\caption{Input and output dimensions for the blocks in the \textsc{Retro} architecture.}
\label{tab:arch_in_out}
\small
\begin{tabular}{lll}
\toprule
{Block}& {Input}& {Output}\\ 
\midrule
\textbf{\begin{tabular}[c]{@{}l@{}}Sentence-\\ Transformer\end{tabular}}& {[}chunk\_len{]}& {[}d\_model{]}\\
\textbf{\begin{tabular}[c]{@{}l@{}}GPT-2 \\ Embedding\end{tabular}}& {[}seq\_len{]}& {[}seq\_len, d\_model{]}\\ 
\textbf{\begin{tabular}[c]{@{}l@{}}GPT-2 \\ Attention\end{tabular}}& {[}seq\_len, d\_model{]}& {[}seq\_len, d\_model{]}\\ 
\textbf{CCA}& \begin{tabular}[c]{@{}l@{}}{[}seq\_len, d\_model{]},\\ {[}n\_chunks, n\_neighbors, neighbor\_len, d\_model{]}\end{tabular}&{[}n\_chunks, n\_neighbors, neighbor\_len, d\_model{]}\\ 
\textbf{FFW}& {[}seq\_len, d\_model{]}& {[}seq\_len, d\_model{]}\\ 
\textbf{READ} & {[}n\_embed{]}   & {[}d\_model, vocab\_size{]}\\ 
\textbf{\begin{tabular}[c]{@{}l@{}}Nearest \\ Neighbor \\ Encoder\end{tabular}} & \begin{tabular}[c]{@{}l@{}}{[}seq\_len, d\_model{]},\\ {[}n\_chunks, n\_neighbors, neighbor\_len, d\_model{]}\end{tabular} & {[}n\_chunks, n\_neighbors, neighbor\_len, d\_model{]} \\ 
\textbf{Normalizer}& {[}n\_chunks, n\_neighbors, neighbor\_len, d\_model{]} & {[}n\_chunks, n\_neighbors, neighbor\_len, d\_model{]} \\ 
\bottomrule
\end{tabular}%
\end{table}

\textbf{Creating the Index}
The workflow begins with the raw text files chosen for the retrieval database. We tokenize the text using the GPT-2 tokenizer. This first tokenization is only used to divide the text into chunks of length 64. To add them to the index, we de-tokenize the chunks, and re-tokenize them using \textsc{Bert} / \textsc{SBert}, so we can embed them using these models. We prefer them over GPT-2 embeddings as they more effectively capture the semantic meaning. These chunk embeddings are then added to the \textsc{Faiss} index as keys, with their continuations (the next 64 tokens) as values.

Before we tokenize with GPT-2, we normalize the text using the \textsc{Bert} normalizer. This is essential, as \textsc{Bert} and GPT treat neither special characters nor repeated characters the same way. Without normalization, the re-tokenized chunk might be longer than 512 tokens. The pseudo-code for this algorithm is in Algorithm \ref{algo:index_creation} in detail.

\begin{algorithm}[H]
\caption{Creating the index.}
\begin{algorithmic}
\State $index \gets IndexIVFPQ()$
\State $X \gets \textsc{Input text}$
\State $X_{norm} \gets bert\_normalizer(X)$
\State $X_{tokenized} \gets GPT2\_tokenizer(X_{norm})$
\State $chunks \gets split\_in\_chunks(X_{tokenized})$
\State $chunks_{text} \gets $[$GPT2\_tokenizer.decode(c)$ for c in $chunks$]
\State $chunks_{emb} \gets $[$SBert\_embedding(c)$ for c in $chunks_{text}$]
\State $index \gets index.add(chunks_{emb})$
\end{algorithmic}
\label{algo:index_creation}
\end{algorithm}
IVFPQ \footnote{https://faiss.ai/cpp\_api/struct/structfaiss\_1\_1IndexIVFPQ.html.} refers to an inverted file with product quantizer. The vectors are clustered and an inverted list for these clusters is created. 
\begin{algorithm}[H]
\caption{Creating the training data.}
\begin{algorithmic}
\State $X \gets \textsc{Input text}$
\State $X_{norm} \gets bert\_normalizer(X)$
\State $X_{tokenized} \gets GPT2\_tokenizer(X_{norm})$
\For{$i = 0; i < len(X_{tokenized}); i = i+1024$}
    \State $src \gets seq[i:i+1024]$
    \State $tgt \gets seq[i+1:i+1025]$
    \State $chunks \gets split\_in\_chunks(src)$
    \For{\texttt{c in chunks}}
    \State $c_{text} \gets GPT2\_tokenizer.decode(c)$
    \State $c_{emb} \gets SBert\_embedding(c)$
    \State $neighbors, distances \gets index.search(c_{emb})$
    \State $save (src, neighbors, distances, tgt)$
    \EndFor
\EndFor
\end{algorithmic}
\label{algo:json_creation}
\end{algorithm}

\textbf{Getting the nearest neighbors}
To make training more efficient, for each training sequence, we get the ten nearest neighbors for each chunk offline. We save them and their associated distance matrices to be loaded during training. In the first step, we again normalize using the \textsc{Bert} normalizer and tokenize using the GPT-2 tokenizer. Then we go through the entire tokenized text in steps of 1024 (our sequence length). Each sequence is split into chunks of length 64, which are used to query the index for their top k neighbors of each chunk. Once found, we save not only the source sequence, the nearest neighbors for all 16 chunks, and their distance matrices but also the target sequence, which is the source sequence shifted by one token.

\label{appendix:changes}
The distance matrices are not strictly necessary for our training setup and can be excluded. In our case, we kept them in order to analyze if adding noise to the distance matrix, thus shuffling the top 10 neighbors, then taking the new top k neighbors would significantly impact the performance. The goal of this experiment was to argue that saving these neighbors on specialized hardware with inherent noise is still feasible. It turned out that our system is even more robust, where we can not only shuffle the neighbors but also add noise to the neighbor embeddings themselves without significant performance degradation.

\subsubsection{Minor Changes}

\textbf{Positional Embeddings}
The positional encodings were ablated in the paper, and Figure 8 shows minimal relevance to the choice of it. Thus, we instead use the state-of-the-art rotary positional embeddings \citep{rope}. They work by using rotational operations to capture positional information. Specifically, they rotate the embeddings based on the position in the sequence. These rotations are learnable and can be reversed.
\\

\textbf{Sequence Length}
We use GPT-2 attention blocks, meaning the input sequence must be of length 1024. In the original paper, the chunk length was 64 tokens and the sequence length was 2048 tokens, neither of which is justified or ablated. As such, to have the input of the correct shape, we reduce the number of chunks per sequence to 16 as opposed to 32, whereas the chunk length remains unchanged.
\\

\textbf{Optimizer and Batch Size}
\textsc{Retro} itself used AdamW \citep{adamw}, so Adam \citep{adam} with true weight decay (as opposed to Adam with L2 regularization), and a linearly increasing learning rate schedule with a batch size of 256 for their smallest model. Due to our memory restrictions, we can work with a batch size of 2 at most. Thus, using the same hyperparameters as Table 11 in \citep{retro} does not yield the same results in terms of convergence. As a consequence, we tried a variety of optimizers and decided on RAdam \citep{radam}, which has the additional benefit of eliminating the need for us to tune the hyperparameters. 

RAdam works by regularizing the variance of the gradients, leading to more stable training, especially in the beginning where the variance might be particularly large. Note that what is displayed as "noam" in the figure, refers to the optimizer used in the original transformer paper \citep{transformer}, where we have Adam with linear warmup and a learning rate schedule that decays the learning rate proportional to the inverse square root of the step number.
\begin{figure}[ht]
    \centering
    \includegraphics[width=0.5\textwidth]{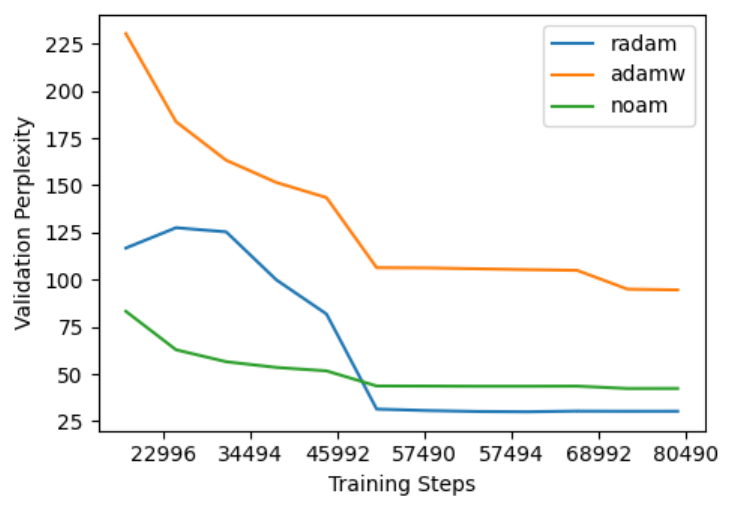}
    \caption{Comparison of various optimizers on WikiText-103.}
    \label{fig:enter-label}
\end{figure}

\subsection{Previous Implementations}
\textsc{Nvidia}'s Megatron project already has a working version of \textsc{Retro}
\citep{nvidiaretro}, but everything from the creation of the search database to the training is based on having much more resources than we do. Moreover, as it is part of a larger project, even following the code flow is difficult, let alone adding changes for this work. Due to these reasons, we chose not to move forward with this implementation of \textsc{Retro}. Although we kept the implementation as a reference to check our architectural decisions against.

As a basis for our experiments, we used the smaller and easier-to-work-with "labml" \citep{labml} implementation of \textsc{Retro} to build our code on. Although only meant as a tutorial to better understand the chunked cross-attention and retrieval mechanism, it has such a clean and simple structure that adding to it for our needs was straightforward. Many changes to this code base were necessary, most importantly to the database, dataset creation, and the training loop, as well as adding \textsc{Retro}-fitted GPT-2. Still, having this framework to start with was hugely beneficial.
\\
\clearpage

\section{Additional Ablations}
Most of our ablations concern the retrieval aspect of the system. Here we explore the language model itself. We wanted to see if it would be possible to fine-tune the GPT-2 attention blocks as well, in order to improve performance even further. It did not, which makes the case for using foundational models stronger.
\label{app:additional_ablations}
\subsection{GPT-2 unfreeze}
\label{app:gpt2_unfreeze}
We kept the GPT-2 backbone frozen for most experiments but did one ablation where we (1) Unfroze GPT-2 and continued to train on one of the pre-trained checkpoints (denoted \emph{from ckp}), (2) Trained from scratch by keeping the GPT-2 parameters trainable (denoted \emph{from scratch}). This increases the number of trainable parameters, which consequently increases the training time.

Considering the results in Table \ref{tab:retro-off-unfreeze-gpt}, we observe that unfreezing GPT-2 and continuing to train from a previous checkpoint does improve our performance slightly for the \textsc{Retro-li}-off case. It is clear, however, that this was due to there no longer being one significant outlier (random seed 42). So it is a stabilizing effect, rather than an overall improved training.

\begin{table}[ht]
    \centering
    \ra{1.3}
    \caption{Perplexity results of \textsc{Retro-li}-off with unfrozen GPT-2 backbone.}
    \vspace{0.5cm}
    \label{tab:retro-off-unfreeze-gpt}
    \begin{subtable}{.5\linewidth}
    \centering
    \caption{Sliding window.}
    \label{tab:retro-off-unfreeze-gpt_75}
    \begin{tabular}{lllll}
        \toprule
        \textsc{Retro-li}-off       &  42             & 43             & 44              & avg\\
        \midrule
        from ckp        &  30.16          &  29.97         & 24.42          & \textbf{28.18}\\
        from scratch    &  \textbf{25.69} &  44.44         &  22.57          & 30.90\\
        frozen GPT      &  55.56          &  \textbf{21.60}&  \textbf{22.02} & 33.06\\
        \bottomrule
    \end{tabular}
    \end{subtable}
    \begin{subtable}{.49\linewidth}
    \centering
    \caption{Full results.}
    \label{tab:retro-off-unfreeze-gpt_100}
    \begin{tabular}{lllll}
    \toprule
        \textsc{Retro-li}-off       &  42             & 43             & 44              & avg\\
        \midrule
        from ckp        &  30.61          &  30.57         &  26.06          & \textbf{29.08}\\
        from scratch    &  \textbf{26.30} &  48.67         &  23.15          & 32.71\\
        frozen GPT      &  53.17          &  \textbf{22.21}&  \textbf{22.59} & 32.66\\
        \bottomrule
    \end{tabular}
    \end{subtable}
\end{table}

Moving on to \textsc{Retro-li}-on (2 neighbors, \textsc{SBert} embeddings) in Table \ref{tab:retro-on-unfreeze-gpt}, the case is even clearer, namely unfreezing GPT-2 did not improve our language modeling capabilities. This is likely because GPT-2 was trained to a minima of language modeling, so when adding retrieval we only have to train the retrieval parameters. Adding chunked cross-attention changed the loss landscape, which is why unfreezing does not help.

\begin{table}[ht]
    \centering
    \caption{Perplexity results of \textsc{Retro-li}-on with unfrozen GPT-2 backbone.}
    \vspace{0.5cm}
    \label{tab:retro-on-unfreeze-gpt}
    \ra{1.3}
    \begin{subtable}{.5\linewidth}
    \centering
    \caption{Sliding window.}
    \begin{tabular}{lllll}
        \toprule
        \textsc{Retro-li}-on        &  42              & 43             & 44              & avg\\
        \midrule
        from ckp        &  26.50           &  27.58         &  27.81          & 27.30\\
        from scratch    &  25.24           &  37.15         &  26.87          & 39.75\\
        frozen GPT      &  \textbf{24.26}  &  \textbf{24.07}&  \textbf{23.82} & \textbf{24.05}\\
        \bottomrule
    \end{tabular}
    \end{subtable}
    \begin{subtable}{.49\linewidth}
    \centering
    \caption{Full results.}
    \begin{tabular}{lllll}
        \toprule
        \textsc{Retro-li}-on         &  42              & 43             & 44              & avg\\
        \midrule
        from ckp        &  27.09           &  28.90         &  28.48          & 28.15\\
        from scratch    &  25.75           &  42.32         &  27.62          & 31.90\\
        frozen GPT      &  \textbf{24.95}  &  \textbf{24.84}&  \textbf{24.39} & \textbf{24.73}\\
        \bottomrule
    \end{tabular}
    \end{subtable}
\end{table}
\clearpage
\newpage
\section{\textsc{Faiss} Index}
\label{app:faiss}
\subsection{Vector indexes}
\label{sec:vector_indexes}
On the non-parametric side, we need to store document vectors in a manner that facilitates search and optimizes space usage.

Maximum inner product search (MIPS) goes through all the vectors in a database, computes the inner product with the query vector, and returns k vectors with the largest inner product. This is linear in the number of entries of the database, thus entirely impractical for databases consisting of millions or even billions of entries. Performing such a search is recommended only when the quality of the retrieved neighbors is crucial and the search time is less critical.

In order to reduce the search time, we must reduce the search scope. This can be done through locality sensitivity hashing (LSH). LSH works by hashing the vectors into b buckets, aiming to group similar vectors based on the hash function. Therefore, it might perform sub-optimally if the vectors are poorly distributed. Moreover, it is highly sensitive to the number of buckets b and the hash function.

We can also reduce the scope by adopting an inverted vector files (IVF) approach, like \textsc{Faiss} does. An IVF index clusters the vectors into c centroids, where each vector is assigned to one centroid. Upon receiving a query, we search $nprobe$ of those c cells to find the nearest neighbors. This reduces the search time while preserving good performance. However, an IVF index requires that we train it in order to identify optimal centroids.

Both index types reduce data transfer by decreasing the search scope. In any case, there is a trade-off between search time and result quality. We have established the importance of good neighbors for training a RAG model. Such approximations would not be necessary if the search could be done in memory. This move to an IMC-based platform would lead to the introduction of noise and non-determinism. However, we have demonstrated that any hardware approximate noise is negligible, especially compared to the approximations made by the index itself.

For illustration purposes, we measure the time it takes for the search function call to \texttt{IndexIVF\_search} to complete and show the results in Table~\ref{fig:faiss_table}. This function is called on 16 chunks concurrently on a Tesla V100 GPU.
\subsection{Profiling Details}
As a way of profiling the speed of the index calls on \textsc{Faiss} we measure the time it takes for the search function call to \texttt{IndexIVF\_search} to complete. This function is called on 16 chunks concurrently. We compare three retrieval databases, WikiText-103 with 1'877'559 entries, CNN-DailyMail with 3'488'000 entries, and SlimPajama with 45'107'000 entries. As the number of database entries almost doubles from WikiText-103 to CNN-DailyMail, so does the average search time. Due to the number of database entries, both have an index with 1024 centroids. However, as the number of database entries becomes 12 times larger from CNN-DailyMail to SlimPajama, the number of centroids becomes 4 times larger and the average search time only becomes 6 times longer. The \texttt{nprobe} parameter of how many out of the closest centroids to search is set to $\sqrt{\texttt{ncentroids}}$. 

This emphasizes the role these parameters play in creating the index. Choosing such hyperparameters carefully can be either beneficial or detrimental to the results and speed of the retrieval. It is important to note that naively setting \texttt{nprobe} to \texttt{ncentroids} would lead to a longer search time than simply searching by brute force. \footnote{https://github.com/facebookresearch/faiss/wiki/Faster-search.}

\begin{figure}[ht]
    \centering
    \begin{subfigure}[h]{0.4\linewidth}
    \includegraphics[width=\linewidth]{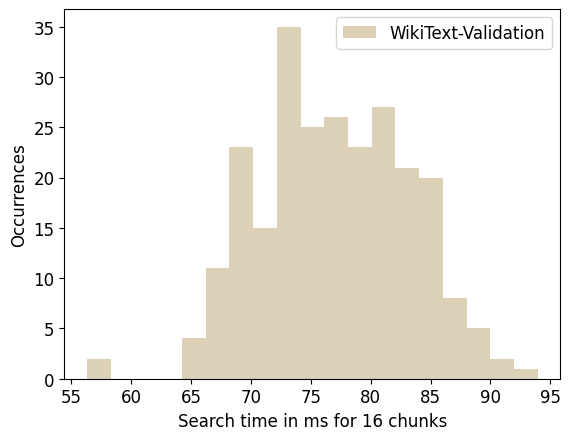}
    \label{fig:faiss_wikitext}
    \end{subfigure}
    \begin{subfigure}[h]{0.4\linewidth}
    \includegraphics[width=\linewidth]{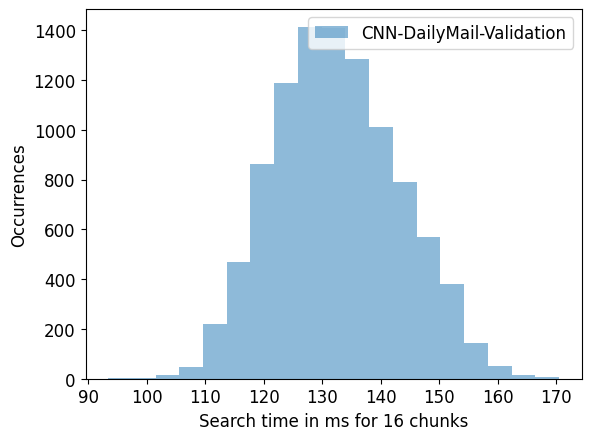}
    \label{fig:faiss_cnndailymail}
    \end{subfigure}
    \begin{subfigure}[h]{0.4\linewidth}
    \includegraphics[width=\linewidth]{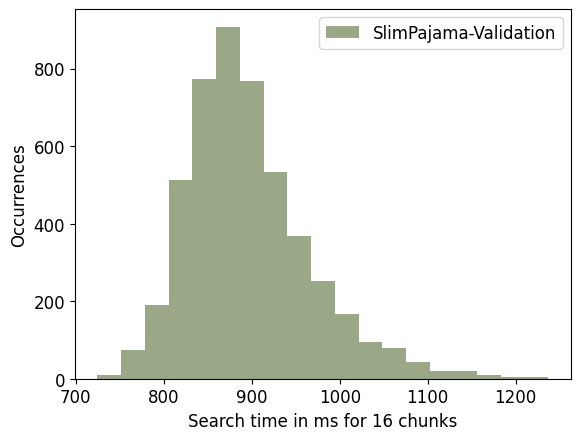}
    \label{fig:faiss_slimpajama}
    \end{subfigure}
    \caption{Histograms of search calls on \textsc{Faiss} index for 16 chunks concurrently, on (a) WikiText-103, (b) CNN-DailyMail, and (c) SlimPajama.}
    \vspace{0.5cm}
    \label{fig:faiss_profiling}
\end{figure}
\clearpage
\newpage
\section{Qualitative Comparison}
\label{app:qualitative_cmp}
Up to this point, all the results have been quantitative. To motivate retrieval, especially in domain shift, we now present some qualitative results. We generate the next chunk based on the context. Generation quality metrics such as fluency, diversity, and repetition depend almost exclusively on how well the generator function is written. \textsc{Nvidia} \citep{nvidiaretro} and DeepMind \citep{retro} addressed this by only taking the greedy output. However, for real use cases, it is common to write the generator function with more care, as simple post-processing steps can make a substantial difference in the fluency and the diversity of the generated output.
\subsection{Multinomial Generation}
The model outputs a $n\_emb \times n\_vocab$ matrix which is interpreted as a probability distribution over the vocabulary words. In greedy generation, we take the vocabulary word with the highest probability, but that can lead to significant degeneration, see Figure \ref{fig:continuation-off-slimpajama-beston}. To generate more natural language, it can be beneficial to occasionally opt for something other than the most likely word.

In multinomial generation, we address this issue with the help of the probability matrix. We cast the matrix as a multinomial distribution and sample from it. This ensures on one hand that it remains likely that we generate words with a higher probability, but also enables us to eventually pick less probable words as well, improving diversity and reducing repetition.
\subsection{Top-p Generation}
Top-p generation is also called nucleus sampling and aims to address similar issues as multinomial generation. It also helps with diversity of language and repetition, but in this case, the focus is not just on reducing the use of common words, but eliminating them altogether. In nucleus sampling, we set a parameter $p$ describing the cumulative probability. We then remove all tokens in the vocabulary with a cumulative probability above this threshold. This results in a new probability distribution, from which we now sample the next token.

\subsection{Results}
We generate the next chunks for the best checkpoints and the best-performing validation samples of \textsc{Retro-li}-on and -off. We chose the datasets WikiText-103 and SlimPajama, which are the two most extreme cases in terms of perplexity. In order to paint a fuller picture, we present the greedy output, the multinomial and top-p generated results, and the real continuation. 

Although there has been tremendous success in evaluating generated samples with GPT-4 (for instance \citep{neftune} used it and motivated its use by comparing it to human judgment) we had no access to it. GPT-3.5-Turbo rankings of the generated samples are unfortunately not deterministic, and even its evaluation of this handful of qualitative examples rarely aligned with human judgment.

\subsubsection{WikiText-103}
For WikiText-103 both \textsc{Retro-li}-on and -off have the same validation sample with the lowest perplexity, so their continuations can be compared directly. Here we anticipate little difference in the outputs between off and on, as the perplexities are close. Their generated outputs are similar in fluency and comprehensiveness.
\begin{figure}[ht]
    \centering
    \includegraphics[width=1\linewidth]{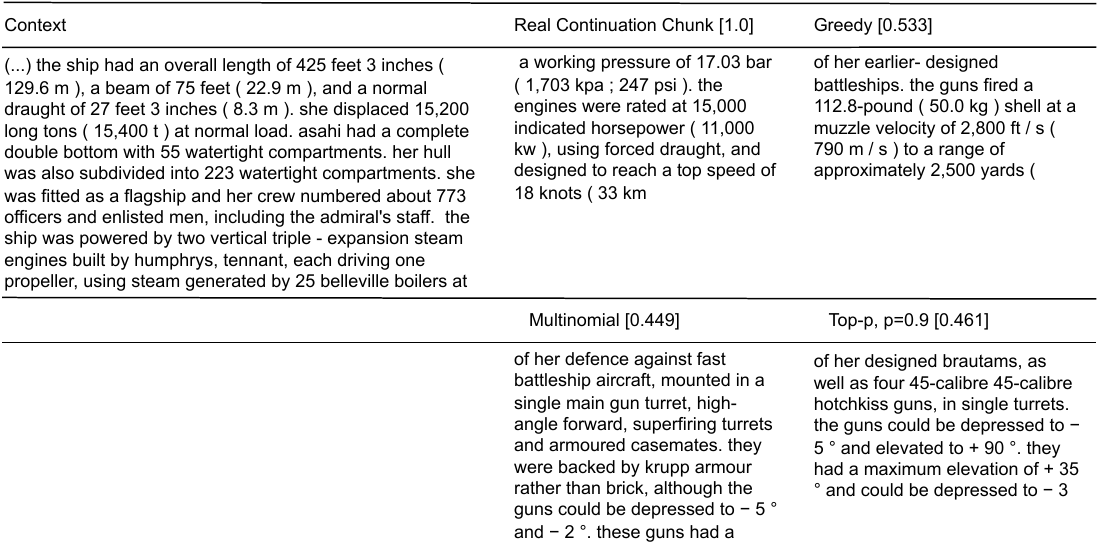}
    \caption{\textsc{Retro-li}-off generated text for WikiText-103-Validation sample with lowest \textsc{Retro-li}-off perplexity.}
    \vspace{0.5cm}
    \label{fig:continuation-off-wikitext}
\end{figure}

\begin{figure}[ht]
    \centering
    \includegraphics[width=\linewidth]{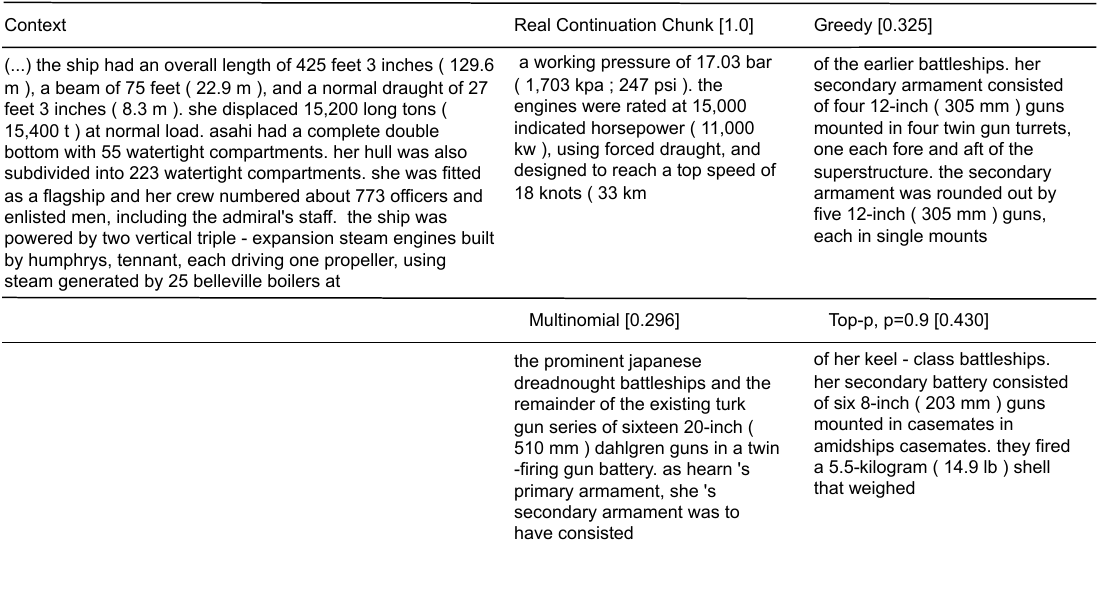}
    \caption{\textsc{Retro-li}-on generated text for WikiText-103-Validation sample with lowest \textsc{Retro-li}-on perplexity.}
    \vspace{0.5cm}
    \label{fig:continuation-on-wikitext}
\end{figure}

\begin{figure}[ht]
    \centering
    \includegraphics[width=\linewidth]{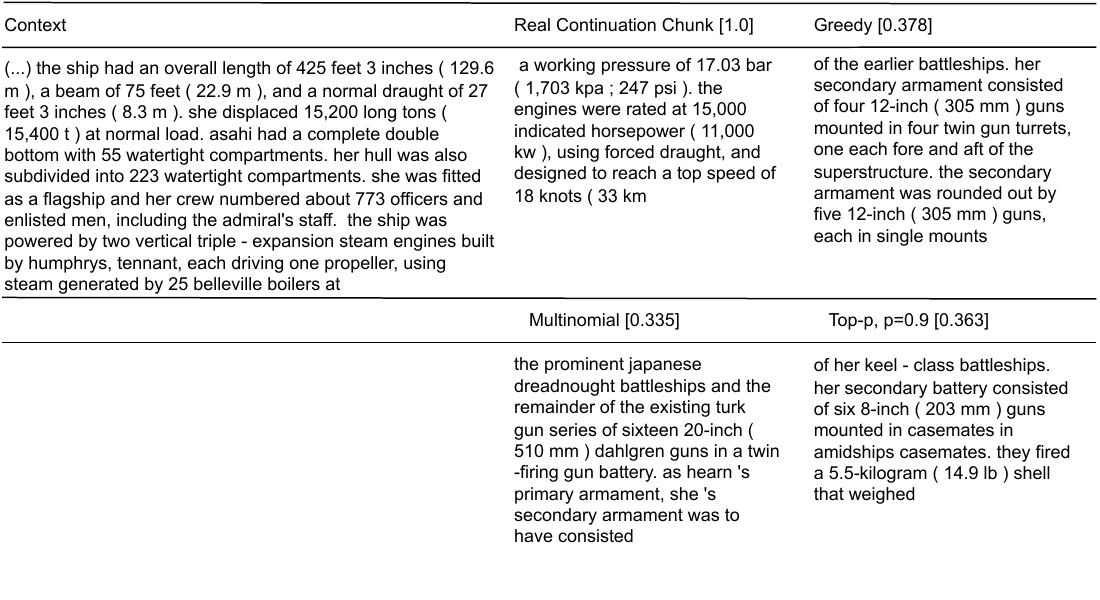}
    \caption{\textsc{Retro-li}-on trained with a Gaussian regularizer with $\lambda_t = 0.2$, generated text for WikiText-103-Validation sample with lowest \textsc{Retro-li}-on perplexity.}
    \vspace{0.5cm}
    \label{fig:continuation-on-reg-wikitext}
\end{figure}

\subsubsection{SlimPajama}
\begin{figure}[ht]
    \centering
    \includegraphics[width=1\linewidth]{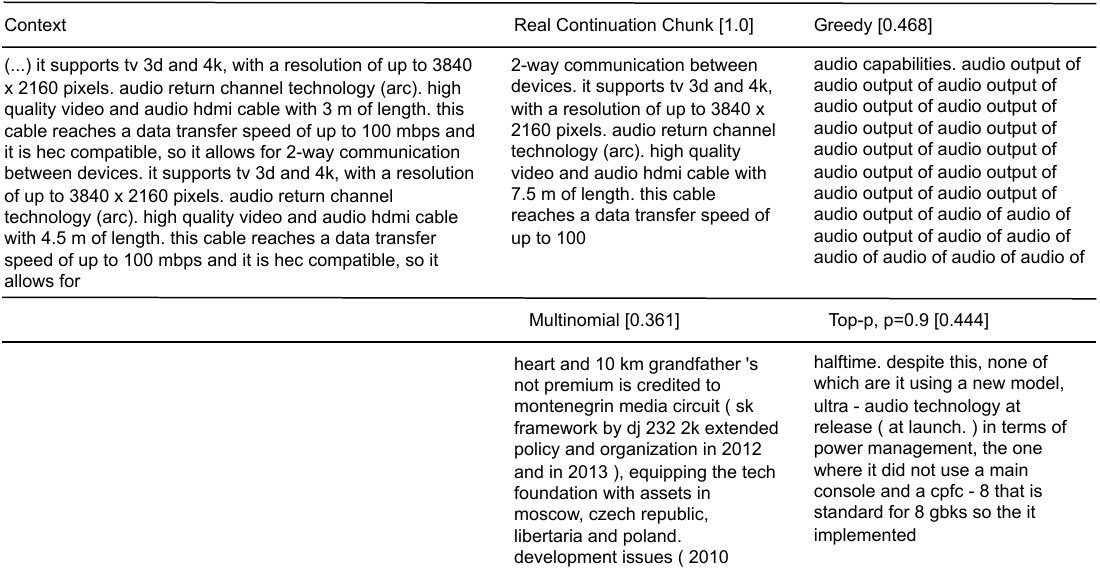}
    \caption{\textsc{Retro-li}-off generated text for the SlimPajama-6B-Validation sample with the lowest \textsc{Retro-li}-on perplexity for comparison purposes.}
    \vspace{0.5cm}
    \label{fig:continuation-off-slimpajama-beston}
\end{figure}

\begin{figure}[ht]
    \centering
    \includegraphics[width=1\linewidth]{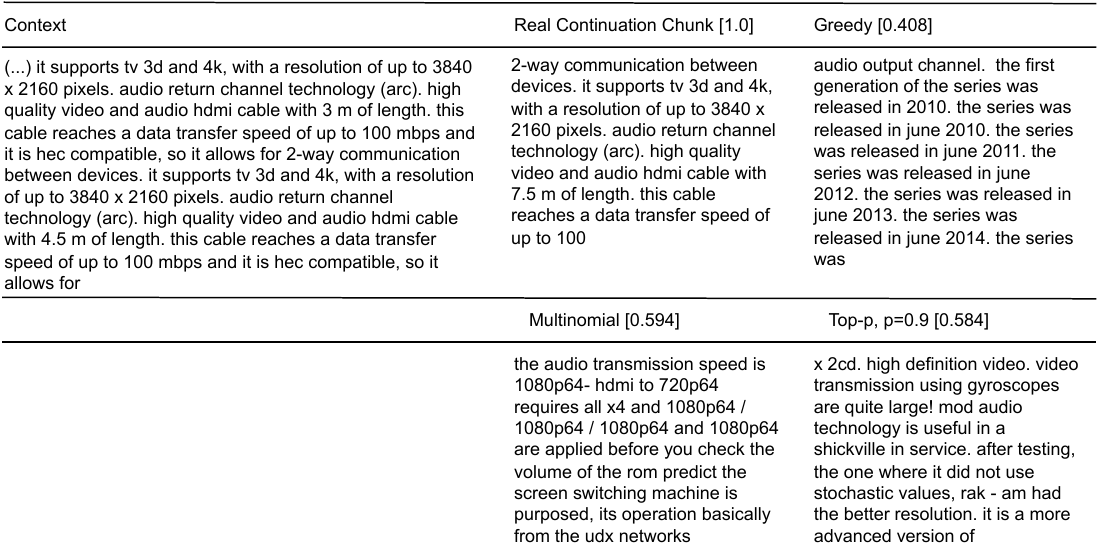}
    \caption{\textsc{Retro-li}-on generated text for the SlimPajama-6B-Validation sample with the lowest \textsc{Retro-li}-on perplexity.}
    \label{fig:continuation-on-slimpajama}
\end{figure}

The SlimPajama dataset is by far the worst in terms of perplexity. Though this can easily be addressed with minimal fine-tuning (see Section \ref{app:finetuning_domainshift}), we are more interested in true \textit{plug-and-play} performance, as it pertains to retrieval. The question becomes: Can retrieval alone help a language model generalize to unseen domains? In Section \ref{sec:results} we show that this is the case quantitatively, here we analyze the output qualitatively as well. Looking at Figure \ref{fig:continuation-off-slimpajama-beston} and Figure \ref{fig:continuation-on-slimpajama} side-by-side, we can see the benefit of retrieval. The topic of this excerpt is a description of an HDMI cable. For \textsc{Retro-li}-off the generated outputs are vaguely about technology whereas \textsc{Retro-li}-on's output is much more on-topic with respect to actual HDMI cables. 

To be completely fair, it must be said that this is the best validation sample for \textsc{Retro-li}-on, not for \textsc{Retro-li}-off.

\begin{figure}[ht]
    \centering
    \includegraphics[width=1\linewidth]{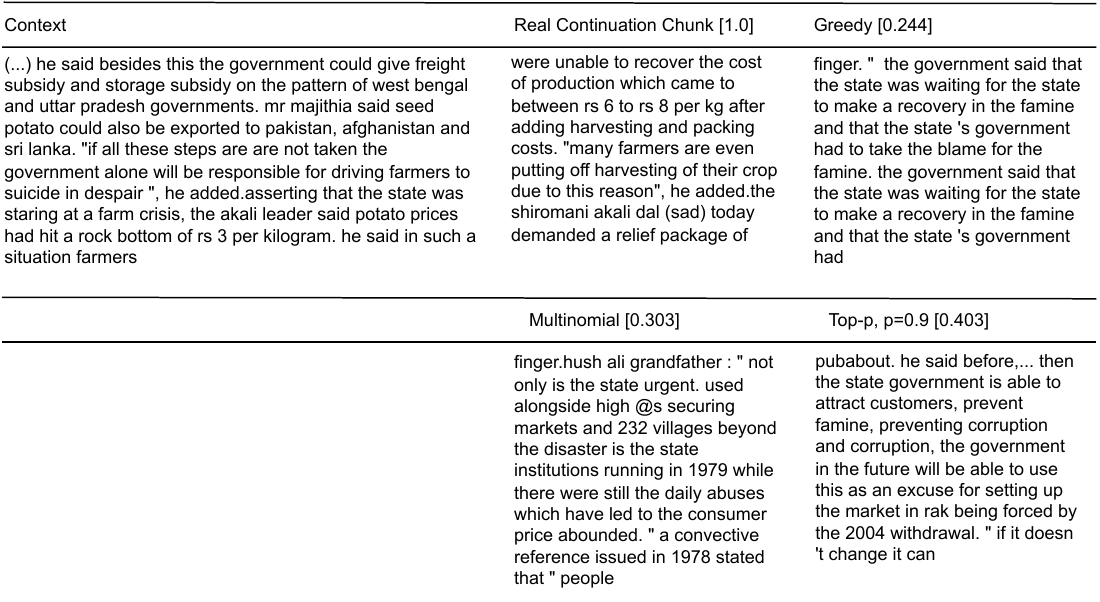}
    \caption{\textsc{Retro-li}-off generated text for the SlimPajama-6B-Validation sample with the lowest \textsc{Retro-li}-off perplexity.}
    \label{fig:continuation-off-slimpajama}
\end{figure}
\clearpage
\subsection{Fine-tuning}
\label{app:finetuning_domainshift}
Although our focus so far has been on plug-and-play performance, fine-tuning is straightforward and improves the perplexities drastically with very few samples. We show some results here for minimal fine-tuning, on just one random seed, for our model trained with a Gaussian regularizer, $\lambda_t=0.2$. We set the number of samples to fine-tune on to 10\% of the validation data size, to provide a balanced assessment.
\begin{table}[ht]
\centering
\caption{Results of fine-tuning on other domains.}
\vspace{0.5cm}
\ra{1.3}
\begin{tabular}{llll}
\toprule
Dataset& \begin{tabular}[c]{@{}l@{}}\# Train \\ samples\end{tabular} & \begin{tabular}[c]{@{}l@{}}Train time\\ {[}min{]}\end{tabular} & \begin{tabular}[c]{@{}l@{}}Validation  \\perplexity\end{tabular}\\ 
\midrule
BBC-News          & 44& 10& 65.89 \\
Reuters           & 17& 4& 54.37 \\
Atticus contracts & 742& 6& 16.57  \\
Founding docs     & 1'816& 6& 48.99  \\
CNN\_DailyMail    & 987& 6& 39.96  \\
OpenWebText       & 1'358& 4& 54.10  \\
SlimPajama        & 483& 5& 82.24 \\
\bottomrule
\end{tabular}
\end{table}

Interestingly, neither the initial perplexity nor the number of samples we fine-tune on is an indicator of how well the dataset takes to fine-tuning.
Overall, we never have to fine-tune for longer than a few minutes to reach a perplexity of around 40.

BBC-News has been fine-tuned the longest and yet has the second-highest final perplexity. This indicates that either the samples are chosen poorly or our model has trouble with this dataset in particular. Atticus contracts, on the other hand, dropped to a perplexity of around 16, which can be attributed to the repetitive nature of contracts. 
\clearpage
\newpage

\end{document}